


\input phyzzx
\message{For printing the figures, you need to have the
files epsf.tex, 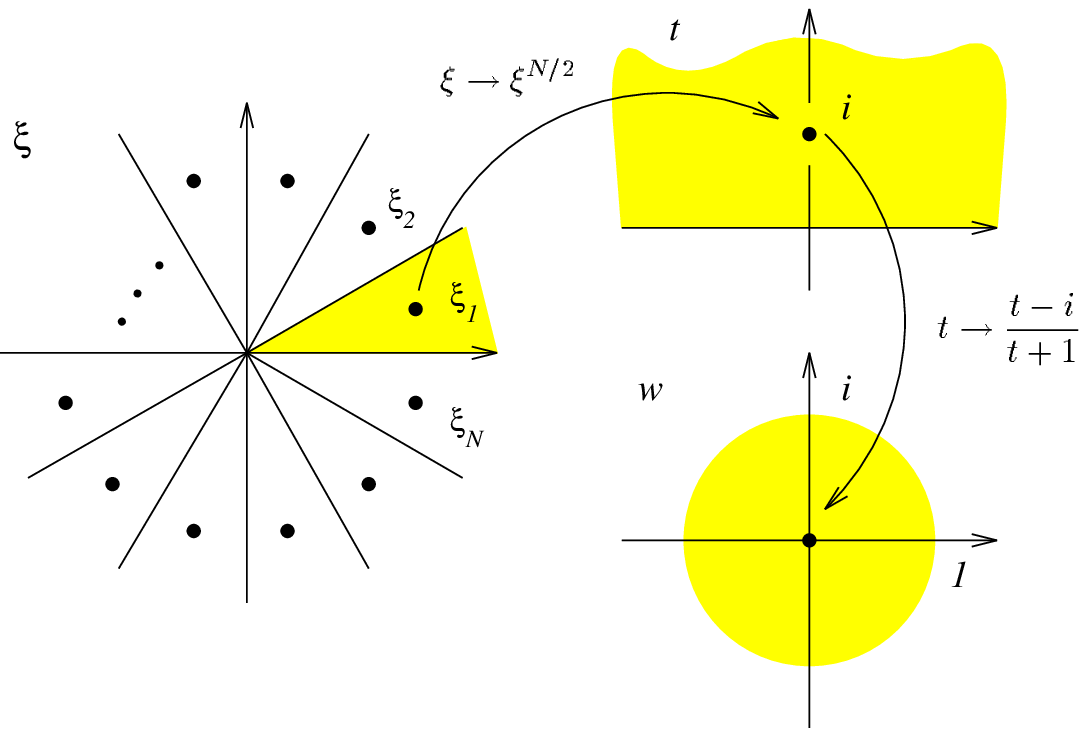 and tanvec.eps
in your area. Do you have these files? Enter y/n :    }
\read -1 to \figcount
\if y\figcount \message{This will come out with the figures}
\else \message{This will come out without the figures}\fi

\if y\figcount
    \input epsf
\else\fi

\hsize=40pc
\catcode`\@=11 
\def\NEWrefmark#1{\step@ver{{\;#1}}}
\catcode`\@=12 

\def\square{\kern1pt\vbox{\hrule height 1.2pt\hbox{\vrule width 1.2pt\hskip 3pt
   \vbox{\vskip 6pt}\hskip 3pt\vrule width 0.6pt}\hrule height 0.6pt}\kern1pt}
\def\A{{\cal A}}
\def\B{{\cal B}}
\def\B{{\cal B}}

\def\D{{\cal D}}
\def\F{{\cal F}}

\def\HH{{\cal H}}
\def\H{\widehat{\cal H}}
\def\I{{\cal I}}

\def\L{{\cal L}}
\def\M{{\cal M}}
\def\N{{\cal N}}
\def\O{{\cal O}}

\def\P{{\cal P}}

\def\V{{\cal V}}
\def\V{{\cal V}}

\def\\#1{_{{}_{#1}}}

\def\bra#1{\langle #1 |}
\def\braket#1#2{\langle#1|#2\rangle}

\def\b{{\bf b}}

\def\ds{\displaystyle}
\def\d{{\rm d}}
\def\e{{\epsilon}}
\def\fact{\left({i\over2\pi}\right)^{N-3}}

\def\frac#1#2{{#1\over#2}}

\def\hP{\widehat\P}
\def\im{\Im{\rm m}}
\def\inbar{\,\vrule height 1.5ex width .4pt depth0pt}\def\IC{\relax
\hbox{$\inbar\kern-.3em{\rm C}$}}
\def\ket#1{| #1 \rangle}
\def\ket#1{| #1 \rangle}

\def\l{\lambda}
\def\ov{\overline}
\def\psl{{\rm PSL}(2,\IC)}
\def\p{\partial}

\def\s{sect.~}

\def\st{S^{{\rm tach}}}
\def\sv{{\bf v}}

\def\v{{\rm v}}

\def\wh{\widehat}


\def\define#1#2\par{\def#1{\Ref#1{#2}\edef#1{\noexpand\refmark{#1}}}}
\def\con#1#2\noc{\let\?=\Ref\let\<=\refmark\let\Ref=\REFS
         \let\refmark=\undefined#1\let\Ref=\REFSCON#2
         \let\Ref=\?\let\refmark=\<\refsend}

\let\refmark=\NEWrefmark

\define\mahapatra{S. Mahapatra and S. Mukherji, `Tachyon Condensates and
Anisotropic Universe', August 1994, hep-th/9408063.}

\define\raiten{E. Raiten, `Tachyon condensates and string theoretic
inflation', Nucl. Phys. {\bf B416} (1994) 881.}

\define\kosteleckysamuel{V. A. Kostelecky and S. Samuel,`On a Nonperturbative
Vacuum for the Open Bosonic String', Nucl. Phys.
{\bf B336} (1990) 263;
`The static Tachyon Potential in the Open Bosonic String Theory',
Phys. Lett. {\bf B207} (1988) 169.}

\define\kosteleckysamueltwo{V. A. Kostelecky and S. Samuel, `Collective
Physics in the Closed Bosonic String, Phys. Rev. {\bf D42} (1990) 1289.}

\define\kosteleckyperry{V. A. Kostelecky and M. J. Perry, `Condensates
and Singularities in String Theory', Nucl. Phys. {\bf B414} (1994) 174,
hep-th/9302120.}

\define\samuelone{S. Samuel, Nucl. Phys.
{\bf B308} (1988) 285; {\bf B308} (1988) 317;
{\bf B323} (1989) 337.}

\define\samueltwo{S. Samuel, Nucl. Phys. {\bf B325} (1989) 275;
{\bf B341} (1990) 513.}

\define\zwiebachlong{B. Zwiebach, `Closed string field theory: Quantum
action and the Batalin-Vilkovisky master equation', Nucl. Phys {\bf B390}
(1993) 33, hep-th/9206084.}

\define\moore{G. Moore, private communication, unpublished.}

\define\banks{T. Banks,`The tachyon potential in String Theory', Nucl. Phys.
{\bf B361} (1991) 166.}

\define\tseytlin{A.~A.~Tseytlin, `On the tachyonic terms in the string
effective action', Phys. Lett {\bf B264} (1991) 311. }

\define\saadizwiebach{M. Saadi and B. Zwiebach, `Closed string
field theory from polyhedra', Ann. Phys. {\bf 192} (1989) 213.}

\define\goldstonethorn{T. L. Curtright, C. B. Thorn and J. Goldstone, `Spin
Content of the Bosonic String', Phys. Lett. {\bf 175B} (1986) 47.}

\define\kugokunitomo{T. Kugo,
H. Kunitomo and K. Suehiro, `Non-polynomial closed string
field theory', Phys. Lett. {\bf 226B} (1989) 48.}

\define\zwiebachma{B. Zwiebach,  `How covariant closed string
theory solves a minimal area problem',
Comm. Math. Phys. {\bf 136} (1991) 83 ; `Consistency of closed string
polyhedra from minimal area', Phys. Lett. {\bf B241} (1990) 343.}

\define\sonodazwiebach{H. Sonoda and B. Zwiebach, `Closed string field theory
loops with symmetric factorizable quadratic differentials',
Nucl. Phys. {\bf B331} (1990) 592.}

\define\polchinski{J. Polchinski, `Factorization of Bosonic String Amplitudes',
Nucl. Phys. {\bf B307}(1988)61.}

\define\nelson{P. Nelson, `Covariant insertion of general vertex operators',
Phys. Rev. Lett.{\bf 62}(1989)993;
H. S. La and P. Nelson, `Effective field equations for fermionic strings',
Nucl. Phys. {\bf B332} (1990) 83;
J. Distler and P. Nelson, `Topological couplings and contact terms in
2-D field theory',
Comm. Math. Phys. {\bf 138} (1991) 273.}

\define\leclair{A. LeClair, M. E. Peskin and C. R. Preitschopf, Nucl.
Phys. {\bf B317} (1989) 411.}

\define\senzwiebach{A. Sen and B. Zwiebach, `Local background independence
of classical closed string field theory',
Nucl. Phys. {\bf B414} (1994) 649,  hep-th/9307088.}

\define\senzwiebachtwo{A. Sen and B. Zwiebach, `Quantum background
independence of closed string field theory', MIT preprint CTP\#2244,
to appear in Nucl. Phys. B, hep-th/9311009.}

\define\senzwiebachnew{A. Sen and B. Zwiebach, `Background Independent
Algebraic Structures in Closed String Field Theory', MIT preprint CTP\#2346,
submitted to Comm. Math. Phys, hep-th/9408053.}

\define\belopolsky{A. Belopolsky, in preparation.}

\define\strebel{K. Strebel,``Quadratic Differentials'', Springer Verlag 1984.}
\overfullrule=0pt
\baselineskip 15pt plus 1pt minus 1pt
\nopubblock

{}~ \hfill \vbox{\hbox{MIT-CTP-2336}
\hbox{hep-th/9409015}\hbox{} }\break
\title{OFF-SHELL CLOSED STRING AMPLITUDES: TOWARDS }
\titlestyle{A COMPUTATION OF THE TACHYON POTENTIAL}

\author{Alexander Belopolsky  \foot{E-mail address: belopols@marie.mit.edu }
and Barton Zwiebach \foot{E-mail address: zwiebach@irene.mit.edu.
\hfill\break Supported in part by D.O.E.
cooperative agreement DE-FC02-94ER40818.}}
\address{Center for Theoretical Physics,\break
Laboratory of Nuclear Science\break
and Department of Physics\break
Massachusetts Institute of Technology\break
Cambridge, Massachusetts 02139, U.S.A.}

\abstract
{We derive an explicit formula for the evaluation of the classical
closed string action for any off-shell string
field, and for the calculation of arbitrary off-shell amplitudes.
The formulae require a parametrization, in terms of some moduli space
coordinates, of the family of local coordinates needed to insert
the off-shell states on Riemann surfaces. We discuss in detail the
evaluation of the tachyon potential as a power series in the tachyon field.
The expansion coefficients in this series are shown to be
geometrical invariants of Strebel quadratic differentials whose
variational properties imply that closed string polyhedra, among
all possible choices of string vertices, yield
a tachyon potential which is as small as possible order by order in the
string coupling constant. Our discussion emphasizes the geometrical
meaning of off-shell amplitudes. }
\endpage

\singlespace
\chapter{Introduction and Summary}

A manifestly background independent field theory of strings should define the
conceptual framework for string theory and should allow the
precise definition and explicit computation of nonperturbative
effects.  The present version of quantum closed string field theory,
developed explicitly only for the case of bosonic strings, while not
manifestly background independent, was proven to be background
independent for the case of nearby backgrounds
[\senzwiebach,\senzwiebachtwo,\senzwiebachnew].
The proof indeed uncovered structures that are expected to be
relevant to the conceptual foundation of string theory.
At the computational level one can ask if present day string field theory
allows one to do new computations, in particular computations that are
not defined in first quantization. While efficient computation may require
the manifestly background independent formulation not yet available, it is
of interest to attempt new computations with present day tools. This is
the main purpose of the present paper.

Off-shell amplitudes are not
naturally defined without a field theory. Indeed, while the basic definition
of an off-shell string amplitude {\it is} given in first quantization,
off-shell
string amplitudes are only interesting if they obey additional properties
such as
permutation symmetry and consistent factorization. These properties are
automatically incorporated when the off-shell amplitudes arise from a
covariant string field theory [\sonodazwiebach].

Off-shell string amplitudes are obtained by integrating
over the relevant moduli space of Riemann surfaces differential forms
that correspond to the correlators of vertex operators inserted at the
punctures of the surfaces and antighost line integrals. The vertex operators
correspond to non-primary fields of the conformal field theory. In contrast,
in on-shell string amplitudes the vertex operators are always
primary fields.
In order to insert  non-primary fields in a punctured Riemann surface we
must choose an analytic local coordinate at every puncture.
The moduli space of Riemann surfaces of genus $g$ and $N$ punctures is
denoted as $\overline\M_{g,N}$, and the moduli space of such surfaces
with choices of local coordinates at the punctures is denoted as
$\wh\P_{g,N}\,$.\foot{The local coordinate
at each puncture is defined only up to a constant phase.}
An off-shell amplitude is just an
integral over a subspace of $\wh\P_{g,N}$. Typically, the relevant
subspaces of $\wh\P_{g,N}$ are sections over $\ov\M_{g,N}$.
Such sections are obtained by making a choice of local coordinates
at every puncture of each surface in $\overline\M_{g,N}$.
In closed string field theory, the use of minimal area metrics
allows one to construct these sections using
the vertices of the theory and the propagator.
Off-shell amplitudes arising in open string field
theory have been studied by Samuel [\samuelone,\samueltwo].

While interesting in their own right, off-shell amplitudes
are not physical observables. More relevant is the evaluation
of the string action for any choice of an off-shell string
field. This computation would be necessary in evaluating string
instanton effects. The string action, apart for the kinetic term, is the
sum of string interactions each of which is defined by a {\it string vertex},
namely, a subspace $\V_{g,N}$ of $\wh\P_{g,N}$. Typically $\V_{g,N}$ is
a section over a compact subspace of $\ov\M_{g,N}$.
Therefore, given an off-shell string field, the contribution to the string
action arising from a particular interaction corresponds to a {\it partially
integrated} off-shell amplitude. The classical potential of a field theory
in flat Minkowski space is a simple example of the above considerations;
it amounts to the evaluation of the action for
field configurations that are spacetime constants.
Ideally we would like to compute, for the case of bosonic strings formulated
around the twenty-six dimensional Minkowski space, the complete classical
potential for the string field. This may be eventually possible but we
address here the computation of the classical potential for some string
modes. In particular we focus in the case of the tachyon of the closed bosonic
string.

For the case of open strings some interesting
results have been obtained concerning the classical {\it effective} potential
for the tachyon [\kosteleckysamuel]. This potential takes into account the
effect of all other fields at the classical level.
In the context of
closed string field theory only the cubic term in the tachyon potential
is known [\kosteleckysamueltwo]. The possible effects of this term have
been considered in Refs.[\kosteleckyperry,\raiten,\mahapatra].
Our interest in the computation of the closed string tachyon potential
was stimulated by G. Moore [\moore]
who derived the following formula for the potential $V(\tau)$ for the
tachyon field $\tau(x)$
$$V (\tau)= -\tau^2-\sum_{n=3}^\infty \,\v\\N\, {\tau^N\over N!}\, ,\quad\quad
{\rm where} \quad \v\\N\, \sim \,\int_{\V_{0,N}} \bigl(\prod_{I=1}^{N-3}
{\d}^2 h\\I\,\bigr) \prod_{I=1}^N
\left|h\\I'(0)\right|^{-2}\, .\eqn\moore$$
This potential is the tachyon potential with all other fields set to zero.
It is not an effective potential. It is fully nonpolynomial, and starts with
a negative sign quadratic term (the symbols appearing in the
expression for $\v\\N$ will be defined in sect.2).
The calculation of the tachyon potential
amounts to the calculation of the constant coefficients $\v\\N$ for
$N\geq 3$. For the cubic term, since $\V_{0,3}$ is
a point, the integral is actually not there, and the evaluation of the
coefficient of $\v_3$ is relatively straightforward.
The higher coefficients are difficult to compute since they involve
integrals over the pieces of moduli spaces $\V_{0,N}$.

We will rewrite \moore\ in
$\psl$ invariant form in order to understand the
geometrical significance of the coefficients $\v\\N$ and to set up
a convenient computation scheme.
Moreover, we will obtain a generalization of eqn.\moore\ valid for
any component
field of the string field theory. The expression will be given in the
operator formalism and will be $\psl$ invariant.

\noindent
\underbar{Extremal Property.}
We will show that the
polyhedral vertices of closed string field theory are the solution to
the problem of minimizing recursively the coefficients in the expansion
defining the tachyon potential. That is, the choice of the Witten vertex,
among all possible choices of cubic string vertices\foot{Vertices are
defined by coordinate curves surrounding the punctures and defining
disks. The disks should not have finite intersection}
minimizes $\v\\3$. Once the three string vertex is chosen, the region
of moduli space to be covered by the four string vertex is fixed. The
choice of the standard tetrahedron for the four string vertex,
among all possible choices of four string vertex filling the required region,
 will minimize the value of $\v\\4$. This continues to be the case for the
complete series defining the classical closed string field theory.
This fact strikes to us as the string field theory doing its best to
obtain a convergent series for the tachyon potential. It is also
interesting that a simple demand, that of minimizing recursively
the coefficients of the tachyon potential, leads uniquely to the polyhedral
vertices of classical closed string field theory. It has been clear
that the consistency of closed string field theory simply depends on having
a choice of string vertices giving a cover of moduli space.
The off-shell behavior, however, is completely
dependent on the choice of vertices,
and one intuitively feels there are choices that are better than other.
We see here nice off-shell behavior arising from polyhedra.

\noindent
\underbar{A Minimum in the Potential?}
In calculating the tachyon potential we must be very careful about sign
factors. The relative signs of the expansion coefficients are essential
to the behavior of the series. We find that all the even terms in the
tachyon potential, including the quadratic term, come with a negative
sign, and all the odd terms come with a positive sign. It then follows,
by a simple sign change in the definition of the tachyon field, that all
the terms in the potential have negative coefficients.
This implies that there is no global minimum in the potential since
the potential is not bounded from below.
Moreover, there is no local minimum that can be identified without detailed
knowledge of the complete  series defining the tachyon
potential.  If the series defining the potential
has no suitable radius of convergence further complications arise in
attempting to extract physical conclusions. We were not able
to settle the issue of convergence, but present some work that goes
in this direction.
In estimating the coefficient $\v\\N$ we must perform an integral of
the tachyon off-shell amplitude over $\V_{0,N}$. In this region the tachyon
amplitude varies strongly. In the middle region the amplitude is lowest, and
if this were the dominating region, we would get convergence. In some
corners of $\V_{0,N}$ the amplitude is so big that, if those
corners dominate, there would be no radius of convergence.

It is important to emphasize that only the tachyon effective potential
(or the full string field potential) is a significant object.
The tachyon potential
is not by itself sufficient to make physical statements. A stable
critical point of this potential may not even be a critical point
of the complete string field potential. The effects of the infinite number
of massive scalar fields must be taken into account.
Our results, making unlikely the existence of a stable critical point
reinforce the  sigma-model arguments that suggest
that bosonic strings do not have time independent stable vacua [\banks],
but are not conclusive.
(See also Ref.[\tseytlin] for a discussion of tachyonic ambiguities in
the sigma model approach to the string effective action.)
The calculation of the full string field potential, or the tachyon effective
potential is clearly desirable. We discuss in Appendix A string field
redefinitions, and argue that it does not seem possible to bring the string
action into a form  (such as one having a purely quadratic tachyon potential)
where one can easily rule out the existence of a local minimum.

\noindent
\underbar{General Off-Shell Amplitudes.}
Since general off-shell computations do not have some of the
simplifying circumstances that are present for the tachyon
(such as being primary, even off-shell), we derive a general formula
useful to compute arbitrary off-shell amplitudes. This formula,
written in the operator formalism, gives the integrand for generic
string amplitudes as a differential form in $\widehat\P_{0,N}$.
The only delicate point here is the construction of the
antighost insertions for Schiffer variations representing
arbitrary families of local coordinates (local coordinates at the
punctures as a function of the position of the punctures on the sphere).
Particular cases of this formula
have appeared in the literature. If the family of local coordinates
happens to arise from a metric, the required antighost insertions were given in
Ref.[\polchinski]. Antighost insertions necessary for zero-momentum dilaton
insertions were calculated in Refs.[\nelson].

\noindent
\underbar{Organization of the Contents of this Paper.}
We now give a brief summary of the contents of the present paper.
In sect.2 we explain what needs to be calculated to extract the
tachyon potential, set up our conventions, and summarize all our
results on the tachyon potential.
In sect.3 we prepare the grounds for the geometrical understanding
of the off-shell amplitudes. We review the definition of the mapping
radius of punctured disks and study its behavior under
$\psl$ transformations (the conformal maps of the Riemann sphere to
itself). We show how to construct $\psl$ invariants for
punctured spheres equipped with coordinate disks, by using
the mapping radii of the punctured disks and coordinate differences between
punctures, both
computed using an arbitrary uniformizer. We review the extremal
properties of Jenkins-Strebel quadratic differentials [\strebel],
and show how our $\psl$ invariants, in addition to having extremal properties,
provide interesting (and seemingly new) functions on the moduli spaces
$\overline\M_{0,N}$.\foot{These functions are analogous to the function
that assigns to an unpunctured Riemann surface the area of the minimal
area metric on that surface.} In sect.4 we compute the off-shell amplitude
for scattering of $N$ tachyons at arbitrary momentum, and give the answer in
terms of integrals of $\psl$ invariants. This formula
is the off-shell extension of the Koba-Nielsen formula. At zero momentum
and partially integrated over moduli space, it gives us, for each $N$, the
coefficient $\v\\N$ of the tachyon potential.  We show why these coefficients
are minimized recursively by the string vertices defined by Strebel
differentials. In sect.5 we do large$-N$ estimates for the coefficients $\v\\N$
of the tachyon potential in an attempt
to establish the existence of a radius of convergence for the series.
The measure of integration is computed exactly for corners in
$\V_{0,N}$ representing a planar configuration for the
tachyon punctures on the sphere. We are also able to estimate
the measure of integration for a uniform distribution of
punctures on the sphere.  In sect.6 we give the operator formalism
construction for general differential forms on $\widehat\P_{0,N}$
labelled by arbitrary off-shell states.

\chapter{String Action and the Tachyon Potential}

In this section we will show what must be calculated in
order to obtain the tachyon potential. This will help put in perspective
the work that will be done in the next few sections.
We will also give some of the necessary conventions, and we will comment
on the significance of the tachyon potential and the limitations of
our results. All of our results concerning the tachyon potential will be
summarized here.

The full string field action is a non-polynomial functional of the
infinite number of fields, and from the component viewpoint, a non-polynomial
function of an infinite number of spacetime fields.
Here we consider only the part of it which contains the tachyon field
$\tau(x)$.
We will call this part the tachyonic action $\st(\tau)$. It is a
nonpolynomial, non-local functional of the tachyonic field $\tau(x)$.
In order to introduce the string field configuration associated to the
tachyon field $\tau(x)$ we first Fourier transform
$$\tau (p) = \int {\d^Dx} \,\, \tau (x)\, e^{-ipx} \, , \eqn\tachft$$
and use $\tau(p)$ to define the tachyon string field $\ket{T}$ as follows
$$\ket{T}=\int{\d^Dp\over (2\pi)^D}\,\tau(p)\, c_1\bar c_1\ket{{\bf 1},p}.
\eqn\tachfield$$
In the conformal field theory representing the bosonic string, the tachyon
vertex operator is given by $T_p = c\bar c e^{ipX}$
and is of conformal dimension
$(L_0 , \overline L_0) =(-1+ p^2/2 , -1 + p^2/2)$. The conformal field theory
state associated to this field is $\,T_p(0)\ket{{\bf 1}} =
c_1\bar c_1 \ket{{\bf 1},p}$. This state is BRST
invariant when we satisfy the on-shell condition
$L_0=\overline L_0 =0$, which requires $p^2=2= -M^2$ (this is the problematic
negative mass squared of the tachyon). The above representative $T_p$ for the
cohomology class of the physical tachyon is particularly nice, because
this tachyon operator remains a primary
field even off-shell ($p^2 \not= 2)$.

The tachyonic action is then given by evaluating the string field action
$S(\ket{\Psi})$ for $\ket{\Psi} = \ket{T}$:
$$\st(\tau)= S\, \left(\, \ket{\Psi} = \ket{T}\, \right) , \eqn\tachact$$
where
$$S(\Psi)={1\over2}\bra{\Psi}c_0^-Q\ket{\Psi}+
     \sum_{N=3}^\infty{\kappa^{N-2}\over N!}\{\Psi^N\}_{\V_{0,N}},\eqn\action$$
and $\kappa$ is the closed string field coupling constant (see
[\zwiebachlong]).
This action
satisfies the classical master equation $ \{S,S\}=0 $ when the string vertices
$\V_0=\sum_{N\geq3}\V_{0,N}$ are chosen to satisfy the recursion relations
$\p\V_0+{1\over2}\{\V_0,\V_0\}=0$ (see Ref.[\senzwiebachnew])

Let us verify that the above definitions lead to the correctly normalized
tachyon kinetic term
$$\st_{{\rm kin}}=\, {1\over2}\,\bra{T}c_0^-Q\ket{T}\,.\eqn\tkin$$
Recall that the BRST operator $Q$ is of the form
$Q=c_0L_0+\bar c_0\bar L_0+\cdots$,
where the dots denote the terms which annihilate $\ket{T}$. Moreover, acting
on the state $c_1\bar c_1 \ket{{\bf 1},p}$ the operators $L_0$
and $\bar L_0$ both have eigenvalue $p^2/2-1$. We then find
$$\st_{{\rm kin}}=
{1\over2}\int{\d^Dp\over (2\pi)^D}\int{\d^Dp'\over (2\pi)^D}
\bra{-p',{\bf 1}}c_{-1}\bar c_{-1}c_0^-c_0^+c_1\bar
c_1\ket{p,{\bf 1}}\,\,\tau(p')\,(p^2-2)\,\tau(p).\eqn\simpta$$
We follow the conventions of Ref.[\zwiebachlong] where
$$\bra{-p',{\bf 1}}c_{-1}\bar c_{-1}c_0^-c_0^+c_1\bar
c_1\ket{p,{\bf 1}}\equiv (2\pi)^D\,\braket{-p',{\bf 1}^c}{p,{\bf
    1}}=(2\pi)^D\delta^D(p'+p)\, ,\eqn\innprod$$
and $c_0^\pm = {1\over 2}(c_0 \pm \bar c_0)$. Using this we finally find
$$\st_{\rm kin}=
-{1\over2}\int{\d^Dp\over (2\pi)^D}\tau(-p)(p^2-2)\tau(p)\, ,\eqn\tchykntc$$
which is indeed the correctly normalized kinetic term.\foot{We work
in euclidean space with positive signature, and the action $S$ should
be inserted in the path integral as $\exp (S/\hbar)$, which is
a convenient convention in string field theory. The euclidean action
$S$ is of the form $S =-\int \d^Dx (K +V)$, where $K$ and $V$ stand for
kinetic and potential terms respectively.}
The $N$-th term in the expansion of the tachyonic action requires the
evaluation of string multilinear functions
$$\st_{0,N}(\tau)={\kappa^{N-2}\over N!}\{T^N\}_{\V_{0,N}}\,,\eqn\termtach$$
and this will be one of the main endeavors in this paper. The answer will be of
the form
$$\{T^N\}_{\V_{0,N}} =
\int \prod\\I {\d p\\I\over (2\pi)^D}\,\, (2\pi)^D
\delta \left( \sum p\\I \right)\cdot V\\N(p_1, \cdots p\\N)\, \tau(p_1)\cdots
\tau(p\\N) \, ,\eqn\fftor$$
where $V$, the function we will be calculating, is well defined
up to terms that vanish upon use of momentum conservation.
To extract from this the tachyon potential we evaluate the above term
in the action for spacetime constant tachyons $\tau(x) = \tau_0$, which
gives $\tau(p) = \tau_0 (2\pi)^D \delta(p)$, and as a consequence
$$\st_{0,N}(\tau_0)= {\kappa^{N-2}\over N!} V_N({\bf 0})\, \tau_0^N
 \cdot \,(2\pi)^D\delta({\bf 0}) \, .\eqn\fftorr$$
Since the infinite $(2\pi)^D\delta({\bf 0})$ factor just corresponds to the
spacetime volume, the tachyon potential will read
$$V (\tau)=  -\tau^2 - \sum_{N\geq 3} {\kappa^{N-2}\over N!} \,\v_{{}\\N}\,\,
\tau^N \, ,\eqn\fftorr$$
where we have used the fact that the potential appears in the action
with a minus sign.
Here the expansion coefficients $\v_{{}\\N}$ are given by
$$\v_{{}\\N}\equiv V\\N({\bf 0})\, . \eqn\littlev$$
We will see that the coefficient $\v\\3$ is given by\foot{The value quoted
here agrees with that quoted in Ref.[\kosteleckyperry] after adjusting
for a factor of two difference in the definition of the dimensionless
coupling constant.}
$$\v_3 = - {3^9\over 2^{11}} \approx - 9.61\,. \eqn\firstermp$$
Analytic work, together with numerical evaluation  gives [\belopolsky]
$$\v_{{}_4} =   72.39\, \pm 0.01 \, .\eqn\numwork$$
Therefore, to this order the tachyon potential reads
$$V(\tau) = -\tau^2 + 1.602 \kappa \tau^3 - 3.016 \kappa^2 \tau^4 +\cdots\,\,,
\eqn\sofartt$$
and gives no local minimum for the tachyon.
The general form for $\v_{{}\\N}$ will be shown to be given by
$$\v_{{}\\N} =(-)^N  {2\over \pi^{N-3}}
\int_{\V_{0,N}} \prod_{I=1}^{N-3}
{\d x\\I d y\\I\over \rho\\I^2}\,\,
{1\over \rho\\{N-2}^2(0) \rho\\{N-1}^2(1) \rho\\N^2(\infty)} \, ,\eqn\coeffs$$
where the quantities $\rho\\I$, called mapping radii, will be discussed
in the next section. Since the integrand is manifestly positive, $\v\\N$
will be positive for even $N$ and negative for odd $N$.
Note that by a sign redefinition of the tachyon field
we can make all terms in the tachyon potential negative.
Therefore the tachyon potential is unbounded from below and
cannot have a global minimum. A local minimum may or
may not exist. Even these statements should be qualified if the
series defining the tachyon potential
has no suitable radius of convergence.
We will study the large-$N$ behavior of the coefficients $\v\\N$ in sect.5,
but we will not be able to reach a definite conclusion as far as the
radius of convergence goes.

Even if one could establish the existence of a local minimum for the
tachyon potential, the question remains whether it represents a vacuum
for the whole string field theory. One way to address this question
would be to compute the effective potential for the tachyon.
For a complete understanding of the string field
potential we should actually examine all zero-momentum Lorentz
scalar fields appearing
in the theory. This would include
physical scalars, unphysical scalars and trivial scalars.
Since even the number of physical scalars
at each mass level grows spectacularly fast [\goldstonethorn],
a more stringy way to discuss the
string field potential is clearly desirable.

\chapter{Geometrical Preliminaries}

In the present section we will begin by reviewing the definition of mapping
radius of a punctured disk. While this object requires a choice of
local coordinate at the puncture, it is possible to use it to construct
conformal invariants of spheres with punctured disks {\it without} having
to make choices of local coordinates at the punctures. We will discuss in
detail those invariants. We review the extremal properties of
the Strebel quadratic differentials and explain how to calculate mapping
radii from them. The invariants relevant to the computation of tachyon
amplitudes are shown to have extremal properties as well.

\section{Reduced Modulus and $\psl$ Invariants}

Given a punctured disk $D$, equipped with a chosen local coordinate $z$
vanishing at the puncture, one can define a conformal invariant
called the mapping radius $\rho_{{}_D}$ of the disk. It is calculated by
mapping conformally the disk $D$ to the unit disk
$|w|\leq 1$, with the puncture going to $w=0$. One then defines
$$ \rho_{{}_D} \equiv \biggl|{dz\over dw}\biggr|_{w=0}\,\,. \eqn\maprad$$
Alternatively one may map $D$ to a round disk $|\xi|\leq \rho\\D$, with
the puncture going to $\xi=0$, so that $|dz/d\xi|_0 = 1$.
The reduced modulus $M_D$ of the disk $D$ is defined to be
$$M_D \equiv  {1\over 2\pi} \ln \rho_{{}_D}\,\,.\eqn\redmod$$
Clearly, both the mapping radius and the reduced modulus depend on the
chosen coordinate. If we change the local coordinate from $z$ to $z'$,
also vanishing at the puncture, we see using \maprad\ that the new
mapping radius ${\rho'}_{{}_D}$ is given by
$${\rho'}_{{}_D} = {\rho}_{{}_D}\, \biggl|{dz'\over dz}\biggr|_{z=0}  \,,
\quad\quad\to \quad {\rho_{{}_D}\over |dz|} = \,{\rm invariant}.\eqn\mrtrans$$
Thus the mapping radius transforms like the inverse of a conformal metric $g$,
for which the length element $g |dz|$ is invariant. For the reduced modulus
we have
$$M'_D=M_D+{1\over 2\pi}\ln\biggl|{dz'\over dz}\biggr|_{z=0}\, .\eqn\mrmtr$$

\noindent
\underbar{$\psl$ Invariants}
It should be noted that the above transformation property \mrtrans\ is not
in contradiction with the conformal invariance of the mapping radius. Conformal
invariance just states that if we map a disk, {\it and}
carry along the chosen local
coordinate at the puncture, the mapping radius does not change. This brings
us to a point that will be quite important. Throughout this paper
we will be dealing with punctured disks on the Riemann sphere. How will we
choose local coordinates at the punctures? It will be done as follows:
we will choose a global uniformizer $z$ on the sphere and keep it fixed.
If a punctured disk $D$ has its puncture at $z=z_0$, then the local
coordinate at the puncture will be taken to be $(z-z_0)$ (the case when
$z_0=\infty$ will be discussed later). Consider now an arbitrary
$\psl$ map taking the sphere into itself
$$z \to f(z) = {az+b\over cz+d}\, ,\eqn\mapsltwoc$$
Under this map a disk $D$ centered at $z=z_0$ will be taken to a disk
$f(D)$ centered at $z= f(z_0)$. According to our conventions, the local
coordinate for $D$ is $z-z_0$ and the local coordinate for $f(D)$ is
$z-f(z_0)$. This latter coordinate is not the image of the original
local coordinate under the map. Therefore the mapping radius
{\it will transform},
and we can use \mrtrans\ to find
$$\rho_{{}_{f(D)}} = \rho_{{}_D} \biggl| {df\over dz}\biggr|_{z_0},\quad\quad
\to \quad \rho_{{}_D} = |cz_0 + d\,|^2 \, \rho_{{}_{f(D)}}\,.\eqn\mrsltwo$$
The off-shell amplitudes will involve the mapping radii of various disks.
Moreover, they must be $\psl$ invariant. How can that be achieved given
that we do not have a $\psl$ invariant definition of the mapping
radius? Let us first examine the case when we have two punctured disks
on a sphere. The data is simply a sphere with two marked points and two
closed Jordan curves each surrounding one of the points. We will associate
a $\psl$ invariant to this sphere. The invariant is calculated
using a uniformizer, but is independent of this choice.
Choose any uniformizer $z$ on the
sphere, and denote the disks by $D_1(z_1)$ and $D_2(z_2)$, where $z_1$ and
$z_2$ are the positions of the punctures. We now claim that
$${\chi_{}}_{ 12} \equiv {|z_1-z_2|^2\over
\rho_{{}_{D_1(z_1)}}\,\rho_{{}_{D_2(z_2)}}} \,, \eqn\twopmi$$
is a $\psl$ invariant (in other words, it is independent of the
uniformizer, or, it is a conformal invariant of the sphere with two punctured
disks). Indeed, under the $\psl$ transformation
given in \mapsltwoc\ we have that
$$ |z_1 - z_2 | = |cz_1 + d\,| \cdot\,| cz_2 + d\, | \cdot
\, | f(z_1) - f(z_2)|\,,\eqn\lftr$$
and it follows immediately from this equation and \mrsltwo\ that
$${|z_1-z_2|^2\over
\rho_{{}_{D_1(z_1)}}\,\rho_{{}_{D_2(z_2)}}}= {|f(z_1)-f(z_2)|^2\over
\rho_{{}_{f( D_1(z_1))}}\,\rho_{{}_{f(D_2(z_2))}}}\,,\eqn\verinv$$
which verifies the claim of invariance of the object $\chi_{12}$. It seems
plausible that any $\psl$ invariant built
from mapping radii of two disks must
be a function of $\chi_{12}$. It is not hard to construct in the same fashion
an $\psl$ invariant of three punctured disks. Indeed, we have
$$\chi_{123} \equiv {|z_1-z_2| \,|z_1-z_3| \, |z_2 -z_3|\over
\rho_{{}_{D_1(z_1)}}\,\rho_{{}_{D_2(z_2)}}\rho_{{}_{D_3(z_3)}} }\,,\eqn\threp$$
which is easily verified to be a conformal invariant. This invariant can
be written in terms of the invariant associated to two disks, one sees that
$$\chi_{123} =  (  \,
\chi_{12}\,\chi_{13}\,\chi_{23}\,)^{1/2}\,. \eqn\notnew$$
This shows the invariant $\chi_{123}$ of three disks is not really new.
It is also clear that we can now construct many invariants of three disks.
We can form linear combinations of complicated functions build using
the invariants associated to all possible choices of two disks from the three
available ones. Nevertheless, the particular invariant $\chi_{123}$ given above
will be of relevance to us later on. Let us finally consider briefly the
case of four punctured disks, and concentrate on invariants having a
product of all mapping radii in the denominator. Let
$$\chi_{1234} \equiv {|z_1-z_2| \,|z_2-z_3| \, |z_3 -z_4|
|z_4 -z_1|\over
\rho_{{}_{D_1(z_1)}}\,\rho_{{}_{D_2(z_2)}}\rho_{{}_{D_3(z_3)}}
\rho_{{}_{D_4(z_4)}}}\,,\eqn\fhrep$$
and, as the reader will have noticed, the only requisite for invariance
is that, as it happens above,
every $z\\I$ appear twice in the numerator. This can be done in
many different ways; for example, we could have written
$$\chi'_{1234} \equiv {|z_1-z_2|^2 \, |z_3 -z_4|^2
\over\rho_{{}_{D_1(z_1)}}\,\rho_{{}_{D_2(z_2)}}\rho_{{}_{D_3(z_3)}}
\rho_{{}_{D_4(z_4)}}}\,,\eqn\fhrepe$$
and the ratio of the two invariants is
$$ {\chi'_4\over \chi_4} = |\lambda|\,,\quad
{\rm with} \quad  \lambda = {(z_1-z_2) \, (z_3 -z_4)\over
(z_1 -z_4)(z_3-z_2) \, }= \{z_1,z_2;z_3,z_4\}\,,  \eqn\crratio$$
which being independent of the mapping radii, and, by construction
a conformal invariant of a four-punctured sphere, necessarily has
to equal the cross-ratio of the four points (or a function of the
cross-ratio). The cross-ratio, as customary, will be denoted by $\lambda$.
It is the point where $z_1$ lands when $z_2,z_3$ and $z_4$ are mapped to
zero, one and infinity, respectively.

\noindent
\underbar{Letting one puncture go to infinity.}
It is sufficient to consider the behavior of the invariant $\chi_{12}$, given
by
$$\chi_{12} \equiv {|z_1-z_2|^2\over
\rho_{{}_{D_1(z_1)}}\,\rho_{{}_{D_2(z_2)}}}\, , \eqn\beginf$$
which we have seen is independent of the chosen uniformizer.
We must examine what happens as we change the uniformizer in such a way
that $z_2 \to \infty$. Given one uniformizer $z$ there is another one
$w=1/z$ that is well defined at $z=\infty$, the only point where
$z$ fails to define a local coordinate. This is why there is no
naive limit to $\chi_2$ as $z_2\to \infty$.
Using \mrtrans\ we express the mapping radius of the second disk in terms
of the mapping radius as viewed using the uniformizer induced by $w$. We have
$\rho_{{}_{D_2(z_2)}}= \Bigl| {dz\over dw}\Bigr|_{w_2}\rho_{{}_{D_2(w_2)}}=
|z_2|^2\rho_{{}_{D_2(w_2)}}$,
and substituting into the expression for $\chi_{12}$ we find
$$\chi_{12} = {|1-z_1/z_2|^2\over
\rho_{{}_{D_1(z_1)}}\,\rho_{{}_{D_2(w_2)}}}\, ,\eqn\tinve$$
and we can now take the limit as $z_2\to \infty$ without difficulty.
Writing, for convenience,
$\rho_{{}_{D_2(w_2=0)}}\equiv\rho_{{}_{D_2(\infty)}}$,
we get
$\chi_{12} = \rho_{{}_{D_1(z_1)}}^{-1}\,\rho_{{}_{D_2(\infty)}}^{-1}\,.$
The apparent dependence of $\chi_{12}$ on the choice of point
$z_1$ is fictitious. Any change
of uniformizer $z\to az+ b$ which changes $z_1$ leaving the point at infinity
fixed, will change the uniformizer at infinity, and the product
of mapping radii will remain
invariant. The point $z_1$ can therefore be chosen to be at the origin, and
we write our final expression for $\chi_2$
$$\chi_{12} = {1\over
\rho_{{}_{D_1(0)}}\,\rho_{{}_{D_2(\infty)}}}\,.\eqn\tiinve$$
Following exactly the same steps with $\chi_3$ and $\chi_4$ we obtain
$$\chi_{123} \equiv {|z_1-z_2| \over
\rho_{{}_{D_1(z_1)}}\,\rho_{{}_{D_2(z_2)}}\rho_{{}_{D_3(\infty)}} }\,,
\eqn\threpp$$
$$\chi_{1234} \equiv {|z_1-z_2| \,|z_2-z_3| \, \over
\rho_{{}_{D_1(z_1)}}\,\rho_{{}_{D_2(z_2)}}\rho_{{}_{D_3(z_3)}}
\rho_{{}_{D_4(\infty)}}}\,.\eqn\fhrepp$$
One could certainly take $z_1=0$ and $z_2=1$ for $\chi_{123}$, and,
$z_2=0$ and $z_3=1$ for $\chi_{1234}$. It should be remembered that whenever
a disk is centered at infinity, the local coordinate used is the inverse of
the chosen uniformizer on the rest of the sphere.

\section{Mapping Radii and Quadratic Differentials}

In this subsection we will review how one uses the Strebel
quadratic differential on a punctured sphere to define punctured disks.
These disks, called coordinate disks, define the local coordinates
used to insert the off-shell states. We will show how one can
use the quadratic differential to calculate the explicit form of the
local coordinates, and the mapping radii of the coordinate disks.
We will review the  extremal properties of the Strebel
quadratic differentials and then discuss the extremal properties of
the $\psl$ invariants.

We will concentrate on the Strebel quadratic differentials relevant for
the restricted polyhedra of closed string field theory. The reader
unfamiliar with these objects may consult Refs.[\strebel,\saadizwiebach].
The Strebel quadratic differential for a sphere with $N$ punctures
in $\V_{0,N}$ induces a metric where the surface can be constructed
by gluing $N$ semiinfinite cylinders of circumference $2\pi$ across
their open boundaries. The gluing pattern is described by a restricted
polyhedron, which is a polyhedron having $N$ faces, each of perimeter
$2\pi$ and, in addition, having all nontrivial closed paths longer than
or equal to $2\pi$. Each semiinfinite cylinder defines a punctured
disk with a local coordinate $w$. The boundary $|w|=1$
corresponds to the edge of the cylinder, to be glued to the polyhedron,
and the puncture corresponds to $w=0$.

The Strebel quadratic differential on the sphere is usually expressed
as $\varphi = \phi(z) (dz)^2$, where $z$ is a uniformizer in the sphere.
At the punctures it has second order poles; if there is a puncture at
$z=z\\I$ the quadratic differential near $z\\I$ reads
$$\varphi = \Bigl( - {1\over (z-z\\I)^2}  + {b_{-1}\over z-z\\I}\, +\, b_0
\,+\, b_1(z-z\\I) \, + \cdots \Bigr)
(dz)^2 \,.\eqn\qdiff$$
Moreover, as mentioned above, the quadratic differential defines a disk
$D\\I$ on the sphere, with the puncture at $z\\I$. A local coordinate $w\\I$
on $D\\I$, such that $D\\I$ becomes a round disk can be found as follows.
We set
$$ z= \rho\\I\, w\\I + c_1w\\I^2 + c_2 w\\I^3 +  \cdots \, , \eqn\ansatz$$
where $\rho\\I, c_1,c_2\cdots$ are constants to be determined. We have
written $\rho\\I$ for the coefficient of $w\\I$ on purpose. If we can
make the $D\\I$ disk correspond to the disk $|w\\I|\leq 1$, then $\rho\\I$
is by definition the mapping radius of the disk $D\\I$, since it is the
value of $|d(z-z\\I)/dw\\I|$ at $w\\I=0$ (recall \maprad). We will actually
use the notation
$$ z = h\\I (w\\I), \quad\to \quad\rho\\I = |h\\I'(0)|\,.\eqn\stanform$$
Note that as explained in the previous subsection we are using the
local uniformizer on the sphere to define the mapping radius.

Back to our problem of defining the $w\\I$ coordinate, we demand that the
quadratic differential, expressed in $w\\I$ coordinates take the form
$$\varphi = - {1\over w\\I^2} (dw\\I)^2 .\eqn\stanform$$
Since the above form is invariant under a change of scale, $w\\I\to aw\\I$,
we cannot determine by this procedure the constant $\rho\\I\,$. If $\rho\\I$
is fixed, the procedure will fix uniquely the higher coefficients
$c_1,c_2,\cdots$. While for general off-shell states the knowledge of the
coefficients $c\\I$ is necessary, for tachyons we only need the mapping
radius. This radius can be determined by the following method. Given
a quadratic differential one must find an arbitrary point $P$ lying
on the boundary of the punctured disk $D\\I$ defined by the quadratic
differential. Possibly, the simplest way to do this is to identify
the zeroes of the quadratic differential and then sketch the critical
trajectories to identify the various punctured disks and ring domains.
One can then pick $P$ to be a zero lying on the nearest
critical trajectory surrounding the puncture.
We now require $w\\I (P)  =1$, and this will fix both the scale and the
phase of the local coordinate. This requirement is satisfied by taking
$$w\\I (z) = \exp \Bigl( i \int_{z(P)}^ z \sqrt{\phi(\xi)}\,  d\xi \, \Bigr)\,,
 \eqn\fixnmormm$$
where we take the positive branch for the square root. If the integral
can be done explicitly then the mapping radius is easily calculated
by taking a derivative $\rho\\I = |{dw\\I\over dz}|_{z\\I}^{-1}$. If the
integral cannot be done explicitly one can calculate the mapping radius
by a limiting procedure. One computes
$\rho\\I = \lim_{\epsilon\to 0} |{dw\\I\over dz}|_{z\\I+\epsilon}^{-1}$.
This leads, using \qdiff\ to the following result
$$\ln \rho\\I = \lim_{\epsilon\to 0} \, \biggl(
\im\hskip-8pt\int_{z\\I+\epsilon}^{z(P)}
\hskip-8pt\sqrt{\phi(\xi)} d\xi\,\,+
\ln \epsilon \,\biggr)\, . \eqn\calcmr$$
The integration path is some curve in the disk $D\\I$, and using contour
deformation one can verify that the imaginary part of the integrand
does not depend on the choice of $P$ as long as $P$ is on the boundary
of $D\\I$. When using equation \calcmr\ one must choose a branch
for the square root, and keep the integration path away from the branch cut.
The sign is fixed by the condition that the limit exist.
Equation \calcmr\ and the recursive procedure
indicated above allow us,
in principle, to calculate the function $h\\I(w\\I)$ if we know explicitly
the quadratic differential.

\noindent
\underbar{Extremal Properties.}
Imagine having an $N$ punctured Riemann sphere and label the punctures
as $P_1,P_2,\cdots P\\N$. Fix completely arbitrary local analytic
coordinates at these punctures. Now consider drawing closed Jordan curves
surrounding the punctures and defining punctured disks $D\\I$, in such
a way that the disks do not overlap (even though they might touch each other).
Given this data we can evaluate the functional
$${\cal F} =  M_{D_1} + M_{D_2} + \cdots + M_{D\\N} \,,\eqn\summod$$
which is simply the sum of the reduced moduli of the various disks.
This functional, of course depends on the shape of the disks we have
chosen, and is well defined since we have picked some specific local
coordinates at the punctures. We may try now to vary the shape of the
disks in order to maximize ${\cal F}$. Suppose there is a choice of
disks that maximizes ${\cal F}$, then, it will maximize ${\cal F}$ whatever
choice of local coordinates we make at the punctures. This follows because
upon change of local coordinates the reduced modulus of a disk changes by a
constant which is independent of the disk itself (see \mrmtr). The interesting
fact is that the Strebel differential defines the disks that maximize
${\cal F}$ [\strebel].
Using the relation between reduced modulus and mapping radius
we see that the functional
$$(\rho_{{}_{D_1}}\cdots\rho_{{}_{D\\N}})^{-1}=\exp(-2\pi{\cal F})\,,\eqn\nfu$$
consisting of the inverse of the product of all the mapping radii, is
actually minimized by the choice of disks made by the Strebel quadratic
differential. This property will be of use to us shortly.

It is worth pausing here to note that the above definition of the functional
${\cal F}$ allows us to compare choices of disks given a {\it fixed} Riemann
sphere. Since we have chosen arbitrarily the local coordinates at the
punctures there is no reasonable way to compare the maximal values of
${\cal F}$ for two {\it different} spheres. It is therefore hard to think
of ${\rm Max}({\cal F})$ as a function on $\overline\M_{0,N}$.
This is reminiscent of the fact
that while for higher genus surfaces {\it without} punctures we can think of
the area of the minimal area metric as a function on moduli space,
it is not clear how to
do this for punctured surfaces. The difficulty again is due to the
regularization needed to render the area finite, this requires a choice
of local coordinates at the punctures, and there is no simple way to
compare the choices for different punctured surfaces. We now wish to
emphasize that our earlier discussion teaches us how
to define functions on $\overline\M_{0,N}$. These functions are
interesting because they are simple modifications of
$\exp[-2\pi \hbox{Max}(\F)]$ that
turn out to be functions on $\overline\M_{0,N}$.

Indeed, consider the invariant $\chi_{1234}$ that was defined as
$$\chi_{1234} \equiv {|z_1-z_2| \,|z_2-z_3| \, |z_3 -z_4|
|z_4 -z_1|\over
\rho_{{}_{D_1(z_1)}}\,\rho_{{}_{D_2(z_2)}}\rho_{{}_{D_3(z_3)}}
\rho_{{}_{D_4(z_4)}}}\,.\eqn\fhrepw$$
Recall that the mapping
radii entering in the definition of $\chi_{1234}$, as well as the
coordinate differences,  are computed using the global uniformizer, and
the invariance of $\chi_{1234}$ just means independence of the result on
the choice of uniformizer. No choice is required to evaluate $\chi_{1234}$.
We obtain a function $f$ on $\overline \M_{0,4}$ by giving a number
for each four punctured sphere $R_4$ as follows. We equip the sphere $R_4$
with the
Strebel quadratic differential $\varphi_S (R_4)$ and we evaluate the invariant
$\chi_{1234}$ using the disks $D^I[\varphi_S (R_4)]$ determined by the
differential. In writing
$$f(R_4) \equiv \chi_{1234}\, \bigl(\, D^I\,[\varphi_S (R_4)]\,\bigr)\, .
\eqn\funcmod$$
We claim that $f(R_4)$ actually is the lowest value that the invariant
$\chi_{1234}$ can take for any choice of nonoverlapping disks in $R_4$
$$\chi_{1234}\, \bigl(\, D^I\,[\varphi_S (R_4)]\,\bigr) \leq
\chi_{1234}\, \bigl(\, D^I\,[ R_4]\,\bigr)  \, .\eqn\claim$$
To see this, fix a uniformizer such that three of the punctures lie at
three points (say, $z=-1,0,1$) and the fourth puncture will lie at some
fixed point, which depends on the choice of four punctured sphere. This
fixes completely the numerator of $\chi$ and fixes the local coordinates
at the punctures, necessary to compute the mapping radii. Therefore
$$\chi_{1234} \propto
(\rho_{{}_{D_1(z_1)}}\,\rho_{{}_{D_2(z_2)}}\rho_{{}_{D_3(z_3)}}
\rho_{{}_{D_4(z_4)}} )^{-1}\, = \exp (-2\pi \F)\,,\eqn\fhrepwx$$
where we recognize that, up to a fixed constant, the invariant is
simply related to the value of $\F$ evaluated with the chosen coordinates
at the punctures. As we now vary the disks around the punctures, $\F$ will
be maximized by the quadratic differential. This verifies that $\chi$
is minimized by the disks chosen by the quadratic differential (Eqn.\claim.)

We expect the function $f(R_4)$
to have a minimum for the most symmetric surface
in $\M_{0,4}$, namely, for the regular tetrahedron
[$\lambda = (1+i\sqrt{3})/2$]. We have not proven this, but
the intuition is that for the most symmetric surface we can get the
disks of largest mapping radii. There is, of course the issue of the
numerator of $\chi$ with the coordinate differences, which also varies
as we move in moduli space. Still, one can convince oneself that
the function $f(R_4)$ grows without bound as $R_4$ approaches degeneration.

\noindent
\underbar{Estimating Mapping Radii.} As we have mentioned earlier,
in defining the mapping radius of a punctured disk on the sphere
we use a local coordinate at the puncture which is obtained from
a chosen uniformizer on the sphere. While this mapping radius depends
on the uniformizer, we are typically interested in functions, such
as the $\chi$ functions, which are constructed out of mapping radii
and coordinate differences, and are independent of the chosen uniformizer.

Consider now the sphere as the complex plane $z$ together with the point
at infinity. The two following facts are useful tools to estimate the
mapping radius of a punctured disk centered at $z_0$.

\noindent
(i) If the disk $D$ is actually a round disk $|z-z_0| \leq R$, then
the mapping radius $\rho_{{}_D}$
is precisely given by the radius of the disk: $\rho_{{}_D} = R$. This
is clear since $w=(z-z_0)/R$ is the exact conformal map of $D$ to a unit disk.

\noindent
(ii) If the disk $D$ is not round but it is contained between two round
disks centered at $z_0$ with radii $R_1$ and $R_2$, with $R_1< R_2$, then
$R_1 < \rho_{{}_D} < R_2$. This property follows from the superadditivity
of the reduced modulus (see Ref.[\zwiebachma] Eqn. (2.2.25)).

Given an $N$ punctured
sphere, the Strebel quadratic differential will maximize the
product of the $N$ mapping radii. We can obtain easily a bound
$\rho_1\rho_2\cdots\rho\\N \geq R_1R_2\cdots R\\N$, where the $R\\I$ are
the radii of non-overlapping round disks centered at the punctures
with the sphere represented as the complex plane together with the point at
infinity.

\chapter{Off-Shell Amplitudes for Tachyons}

In this section we compute off-shell amplitudes for tachyons at
arbitrary momentum. We first discuss the case of three tachyons and
then the case of $N\geq 4$ tachyons which requires integration
over moduli space. We examine the results for the case of zero-momentum
tachyons obtaining in this way the coefficients
$\v\\N$ of the tachyon potential.  We explain why the choice of polyhedra
for the string
vertices, minimizes recursively the coefficients of the nonpolynomial
tachyon potential.

\section{Three Point Couplings}

We will now examine the cubic term in the string field potential.
Assume we are now given a three punctured sphere, and we want to
calculate the general off-shell amplitude for three tachyons. We
then must compute the correlator
$$A_{p_1p_2p_3}=
\Bigl\langle c\bar c e^{ip_1X} (w_1=0)\,\,c\bar c e^{ip_2X} (w_2=0)\,\,
c\bar c e^{ip_3X} (w_3=0) \Bigr\rangle\, .\eqn\correlcx$$
We must transform these operators from the local coordinates $w\\I$
to some uniformizer $z$. Let $w\\I=0$ correspond to $z=z\\I$. We then have
from the transformation law of a primary field
$$ c\bar c e^{ip\\IX} (w\\I=0) =  c\bar c e^{ip\\IX} (z=z\\I)\Bigl|
{dz\over dw\\I}
\Bigr|^{p\\I^2-2}_{w\\I=0}=c\bar ce^{ip\\IX} (z\\I)\,\,
\rho_1^{p\\I^2-2}\,,\eqn\trtach$$
where in the last step we have recognized the appearance of the mapping
radius for the disk $D\\I$. The correlator then becomes
$$\eqalign{
A_{p_1p_2p_3}& =
\Bigl\langle c\bar c e^{ipX} (z_1)\,\,c\bar c e^{ipX} (z_2)\,\,
c\bar c e^{ipX} (z_3) \Bigr\rangle\,{1\over
\rho_1^{2-p_1^2} \rho_2^{2-p_2^2}\rho_3^{2-p_3^2}  }\, ,\cr
&={ |z_1-z_2|^{2+2p_1p_2}|z_2-z_3|^{2+2p_2p_3}|z_1-z_3|^{2+2p_1p_3}\over
\rho_1^{2-p_1^2} \rho_2^{2-p_2^2}\rho_3^{2-p_3^2} \,  }
\cdot [-2(2\pi)^D \delta^D (\sum p\\I)\,]\,,\cr} \eqn\correlc$$
where we made use of \innprod, which introduces an extra factor of
$-2$ (shown in brackets) due to our convention
$c_0^\pm = (c_0 \pm \bar c_0)/2$.
In order to construct a manifestly $\psl$ invariant expression we
use momentum conservation in the denominator to write
$$\eqalign{
A_{p_1p_2p_3}&={ |z_1-z_2|^{2+2p_1p_2}\over (\rho_1\rho_2)^{1+p_1p_2}} \,
{|z_2-z_3|^{2+2p_2p_3}\over (\rho_2\rho_3)^{1+p_2p_3}}\,
{|z_1-z_3|^{2+2p_1p_3}\over
 (\rho_1\rho_3)^{1+p_1p_3}}\cdot  [-2(2\pi)^D \delta^D (\sum p\\I)\,]\,,\cr
 &=  [-2(2\pi)^D \delta^D (\sum p\\I)\,]\cdot
\prod_{I<J}^3 \,[\chi\\{IJ}]^{1+p\\Ip\\J}\, ,\cr}\eqn\minvf$$
which is the manifestly $\psl$ invariant description of the off-shell
amplitude.

We can now use the above result to extract the cubic coefficient of the
tachyon potential. By definition, the operator formalism bra
$\bra{V^{(3)}_{123}}$ representing
the three punctured sphere must satisfy
$$\bra{V^{(3)}_{123}} \left( c_1\bar c_1 \ket{{\bf 1},p_1}\right)^{(1)}
 \left( c_1\bar c_1 \ket{{\bf 1},p_2}\right)^{(2)}
 \left( c_1\bar c_1 \ket{{\bf 1},p_3}\right)^{(3)}\,
= A_{p_1p_2p_3} \,.\eqn\connect$$
Moreover, the multilinear function representing the cubic interaction
is given by
$$\eqalign{
\{ T\}_{\V_{0,3}} &\equiv \bra{V^{(3)}_{123}} T\rangle^{(1)} \ket{T}^{(2)}
\ket{T}^{(3)} \,,\cr
&= \int {\d p_1\over (2\pi)^D}{\d p_2\over (2\pi)^D}{\d p_3\over (2\pi)^D}
\,A_{p_1p_2p_3}\, \tau(p_1) \tau(p_2)\tau(p_3)\, ,\cr
&=\int \prod_{I=1}^3{\d p\\I\over (2\pi)^D}\,
 (2\pi)^D \delta^D (\sum p\\I)\,
\cdot (-2)\prod_{I<J}^3 \,[\chi\\{IJ}]^{1+p\\Ip\\J}\,\cdot
\tau(p_1) \tau(p_2)\tau(p_3) \,,\cr} \eqn\seehow$$
where use was made of the definition of the string tachyon field in
\tachfield, of \connect, and of \minvf.
Comparison with \fftor, and use of \littlev\ now gives
$$\v\\3  = -2\cdot \prod_{I<J}^3 \,[\chi\\{IJ}] = -2\cdot
{ |z_1-z_2|^2|z_2-z_3|^2|z_1-z_3|^2\over
\rho_1^2 \rho_2^2 \rho_3^2}
= -2\cdot \chi_{123}^2\, , \eqn\tprjk$$
in terms of the $\psl$ invariant $\chi_{123}$. It follows from the extremal
properties discussed earlier that the minimum value possible for $\v\\3$
is achieved for the Strebel quadratic differential defining the Witten
vertex. We will calculate the minimum possible value for $\v\\3$ in
sect.5.1.

\section{The Off-Shell Koba-Nielsen Formula}

We now derive a formula for the off-shell scattering amplitude
for $N$ closed string tachyons at arbitrary momentum. The final
result will be a manifestly $\psl$ invariant expression.
The computation is simplified because the tachyon vertex operator
is primary even off-shell, and because its ghost structure is
essentially trivial.

We work in the $z$-plane and fix the position of the three last
insertions at $z\\{N-2}$, $z\\{N-1}$ and $z\\N$. The positions of the
first $N-3$ punctures will be denoted as $z_1,z_2,\cdots z\\{N-3}$.
We must integrate over the positions of these $N-3$ punctures.
Each will therefore
give a factor
$$\d x\\I\wedge \d y\\I\,\, b({\partial\over \partial x})
\,b({\partial\over \partial y}) = \d x\\I\wedge \d y\\I\,
\,\, 2i\,\bar b_{-1} b_{-1} = - \d z\\I\wedge \d\bar z\\I\, \bar b_{-1} b_{-1}
\,, \eqn\fmea$$
where $z\\I = x\\I + iy\\I$.
There is a subtlety here, each of the antighost oscillators refers to
the $z$ plane, while the ghost oscillators in each tachyon insertion
$c_1^{w\\I}\bar c_1^{w\\I}\ket{0,p}^{w\\I}$ refer to the local coordinate
$w\\I$, where
$z= h\\I(w\\I)$. Transforming the antighost oscillators we
obtain $b_{-1} = [h\\I'(0)]^{-1} b^{w\\I}_{-1}+\cdots$, where the dots indicate
antighost oscillators $b_{n\geq 0}^{w\\I}$ that annihilate the tachyon state.
For the antiholomorphic oscillator we have
$\bar b_{-1} = \left[\overline{h\\I'(0)}\right]^{~-1}\bar
b^{w\\I}_{-1}+\cdots$.
Therefore
each of the integrals will be represented by
$$2i\,\d x\\I\wedge \d y\\I\,\,{1\over \rho\\I^2}\,\ket{0,p}^{w\\I}=
2i\,\d x\\I\wedge \d y\\I\,{1\over \rho\\I^{2-p\\I^2}}\,\ket{0,p}\,,\eqn\mfrn$$
where $\rho\\I=|h\\I'(0)|$ is the mapping radius of the $I$-th disk.
The Koba-Nielsen amplitude will therefore be given by
$$\eqalign{
A_{p_1\cdots p_N} & = \left({i\over 2\pi}\right)^{N-3} (2i)^{N-3}
\int \prod_{I=1}^{N-3}
{\d x\\I\d y\\I\,\over \rho\\I^{2-p\\I^2} }
{1\over \rho\\{N-2}^{2-p\\{N-2}^2}
\rho\\{N-1}^{2-p\\{N-1}^2}\rho\\{N-2}^{2-p\\{N-2}^2}}\cr
&\quad \cdot \Bigl\langle
e^{ip_1X(z_1)}\cdots
e^{ip\\{N-3}X(z\\{N-3})}\,\,c\bar ce^{ip\\{N-2}X(z\\{N-2})}
\,c\bar ce^{ip\\{N-1}X(z\\{N-1})}\,
c\bar ce^{ip\\{N}X(z\\{N})}\Bigr\rangle \, ,\cr}\eqn\knoff$$
where the correlator is a free-field correlator in the complex plane.
We will not include in the amplitude the coupling constant factor
$\kappa^{N-3}$.
The extra factor $(i/2\pi)^{N-3}$ included in the formula above
is well-known to be necessary for
consistent factorization, and has been derived in closed string field theory.
\foot{The value used here appears in Ref.[\senzwiebachtwo], where
a sign mistake of Ref. [\zwiebachlong] was corrected.}
We then have
$$\eqalign{
A_{p_1\cdots p_N}  &= (-)^N \cdot
{2\over \pi^{N-3}}
\int \prod_{I=1}^{N-3}
{\d x\\I\d y\\I\over \rho\\I^{2-p\\I^2} }\,\,
{|(z\\{N-2}-z\\{N-1})\,(z\\{N-2}-z\\N)\,(z\\{N-1}-z\\N)|^2
\over \rho\\{N-2}^{2-p\\{N-2}^2}
\rho\\{N-1}^{2-p\\{N-1}^2}\rho\\{N}^{2-p\\{N}^2}}\cr
&\quad \cdot \prod_{I<J}^N |z\\I-z\\J|^{2p\\Ip\\J}
\,\cdot \left[ (2\pi)^D \delta \left( \sum p\\I \right) \right] \,.\cr}
\eqn\knofff$$
Using momentum conservation, and the definition of the invariants
$\chi\\{IJ}$ and $\chi\\{IJK}$, we can write the above as
$$A_{p_1\cdots p_N}  =(-)^N
{2\over \pi^{N-3}}\int \prod_{I=1}^{N-3}
{\d x\\I \d y\\I\over \rho\\I^2}\,\,\chi\\{N-2,N-1,N}^2
\, \cdot \prod_{I<J}^N \chi\\{IJ}^{p\\Ip\\J}\,
\,\cdot \left[ (2\pi)^D \delta \left( \sum p\\I \right)
\right] , \eqn\knofffx$$
It follows immediately from the transformation law for the mapping radius
that the measure $dz\\I\wedge d\bar z\\I / \rho\\I^2$ is $\psl$ invariant.
Therefore the above result is a manifestly $\psl$ invariant off-shell
generalization of the Koba-Nielsen formula.
For the case of four tachyons it reduces to an off-shell version of the
Virasoro-Shapiro amplitude
$$A_{p_1\cdots p_4}  = {2  \over \pi}\int
{\d x_1 \d y_1\over \rho_1^2}\,\,
\chi_{234}^2
\, \cdot \prod_{I<J}^4 \chi\\{IJ}^{p\\Ip\\J}
\,\cdot \left[ (2\pi)^D \delta \left( \sum p\\I
\right) \right] \,.\eqn\knofffx$$
If we choose to place the second, third, and fourth punctures at zero, one
and infinity respectively, we end with
$$A_{p_1\cdots p_4}  = {2  \over \pi}\int
\d x\, \d y \,{ |z|^{2p_1p_2} |z-1|^{2p_1p_3} \over \rho_1^{2-p_1^2}
\rho_2^{2-p_2^2}\rho_3^{2-p_3^2}\rho_4^{2-p_4^2}}
\,\cdot \left[ (2\pi)^D \delta \left( \sum p\\I \right) \right] \,.\eqn\fxyu$$

Another expression can be found where the variables of integration
are cross-ratios. We define
the cross ratio
$$\lambda\\I \equiv \{z\\I, z\\{N-2};z\\{N-1},z\\N\} = {(
z\\I-z\\{N-2})\, (z\\{N-1}-z\\N)\over
\,(z\\I-z\\N)\,\, (z\\{N-1}-z\\{N-2}) }\,,\eqn\cratio$$
and it follows that
$$\d z\\I\wedge \d\bar z\\I = {\d\lambda\\I\wedge \d\bar \lambda\\I
\over |\lambda\\I|^2}\,
{ |z\\I-z\\N|^2 |z\\I-z\\{N-2}|^2\over |z\\N-z\\{N-2}|^2 }\,, \eqn\sdd$$
leading to
$$\eqalign{
A_{p_1\cdots p_N}  &= 2(-)^{N}\left( {i\over 2\pi} \right)^{N-3}
\int \prod_{I=1}^{N-3}
\Bigl[ {\d\lambda\\I\wedge \d\bar \lambda\\I\over |\lambda\\I|^2}
\Bigl( {\chi\\{I,N-2,N-1,N}
\over \chi\\{N-2,N-1,N} }\Bigr)^2 \Bigr]
\, \cdot \chi^2\\{N-2,N-1,N}\cr
&\quad\cdot \prod_{I<J}^N \chi\\{IJ}^{p\\Ip\\J}
\,\cdot \left[ (2\pi)^D \delta \left( \sum p\\I \right) \right] \,. \cr}
\eqn\kknofff$$
For the case of four tachyons the above result reduces to
$$A_{p_1\cdots p_4}  = {i  \over \pi}\int
 {\d\lambda \wedge\d\bar \lambda\over |\lambda|^2}\, \chi_{1234}^2
\,\cdot \prod_{I<J}^4 \chi\\{IJ}^{p\\Ip\\J}
\,\cdot \left[ (2\pi)^D \delta \left( \sum p\\I \right) \right] \,.
\eqn\kknofff$$
In the above expressions the $\lambda$ integrals extend over the whole sphere.

Having obtained manifestly $\psl$ invariant expressions valid for arbitrary
momenta, we now go back to our particular case of interest, which is the
case when all the momenta are zero. It is simplest to go back to \knoff\
to obtain
$$A_{1\cdots N}  = (-)^N {2\over \pi^{N-3}}
\int \prod_{I=1}^{N-3}
{\d x\\I \d y\\I\over \rho\\I^2}\,\,
{|z\\{N-2}-z\\{N-1}|^2|z\\{N-2}-z\\N|^2 |z\\{N-1}-z\\N|^2
\over \rho\\{N-2}^2 \rho\\{N-1}^2 \rho\\N^2}
[(2\pi)^D \delta ({\bf 0})]\,,\eqn\wknoofffx$$
$$A_{1\cdots 4}  = {2\over \pi}\int
{\d x_1\, \d y_1\over \rho_1^2}\,\,
{|z_{2}-z_{3}|^2|z_{2}-z_4|^2 |z_{3}-z_4|^2
\over \rho_{2}^2 \rho_{3}^2 \rho_4^2}
[(2\pi)^D \delta ({\bf 0})]\,,\eqn\wknoqfffx$$
These are the expressions we shall be trying to estimate. If we set the
three special points appearing in the above expressions to zero, one and
infinity, we find (see \s3.3)
$$A_{1\cdots N}  = (-)^N  {2\over \pi^{N-3}}
\int \prod_{I=1}^{N-3}
{\d x\\I d y\\I\over \rho\\I^2}\,\,
{1\over \rho\\{N-2}^2(0) \rho\\{N-1}^2(1) \rho\\N^2(\infty)}
\,[(2\pi)^D \delta ({\bf 0})]\,,\eqn\wfx$$
$$A_{1\cdots 4}  = {2 \over \pi}\int
{\d x_1\, \d y_1\over \rho_1^2}\,\,
{1\over \rho_{2}^2(0) \rho_{3}^2(1) \rho_4^2(\infty)}
\, [(2\pi)^D \delta ({\bf 0})]\, .\eqn\wkx$$

Let us now use the above results to extract the quartic and higher order
coefficient of the tachyon potential.
By definition, the operator formalism bra representing
the collection of $N$-punctured spheres must satisfy
$$\int_{\V_{0,N}}  \bra{\Omega_N^{(0)0,N}}
\left( c_1\bar c_1 \ket{{\bf 1},p_1}\right)^{(1)}
\cdots \left( c_1\bar c_1 \ket{{\bf 1},p_N}\right)^{(N)}\,
= A_{p_1\cdots p_N}(\V_{0,N}) \,,\eqn\conct$$
where the $\V_{0,N}$ argument of $A_{p_1\cdots p_N}(\V_{0,N})$ indicates
that the off-shell amplitude has only been partially integrated over the
subspace $\V_{0,N}$.
The corresponding multilinear function
is given by
$$\{ T\}_{\V_{0,N}} \equiv \int_{\V_{0,N}}  \bra{\Omega_N^{(0)0,N}}
 T\rangle^{(1)}\cdots
\ket{T}^{(N)} \,
= \int \prod_{I=1}^N{\d p\\I\over (2\pi)^D}\,
\,A_{p_1\cdots p_N}(\V_{0,N})\, \tau(p_1) \cdots\tau(p_N)\,, \eqn\seehow$$
where we made use of the definition of the string tachyon field in
\tachfield,  and of \conct.
Reading the value of the amplitude at zero momentum, and by virtue
of \fftor\ and \littlev\ we get
$$\eqalign{
\v\\N  &=  (-)^N{2\over \pi^{N-3}}
\int_{\V_{0,N}} \prod_{I=1}^{N-3}
{\d x\\I \d y\\I\over \rho\\I^2}\,\,
{|z\\{N-2}-z\\{N-1}|^2|z\\{N-2}-z\\N|^2 |z\\{N-1}-z\\N|^2
\over \rho\\{N-2}^2 \rho\\{N-1}^2 \rho\\N^2 }\, , \cr
& = (-)^N {2\over \pi^{N-3}}
\int_{\V_{0,N}} \prod_{I=1}^{N-3}
{\d x\\I d y\\I\over \rho\\I^2}\,\,
{1\over \rho\\{N-2}^2(0) \rho\\{N-1}^2(1) \rho\\N^2(\infty)}\,.\cr}\eqn\trjk$$
Note here the pattern of signs. All $\v\\N$ for $N$ even come with positive
sign, and all $\v\\N$ for $N$ odd come with a negative sign. Including
the overall minus sign in passing from the action to the potential
(Eqn.\fftorr), and the
sign redefinition $\tau \to -\tau$, all coefficients
of the tachyon potential become negative.

It is worthwhile to pause and reflect about the above pattern of signs.
In particular, since $\v\\4$ turned out to be positive, the quartic
term in the tachyon potential is negative, as the quadratic term is.
While the calculations leading to the sign factors are quite subtle,
we believe that the result should have been expected. In closed
string field theory, the elementary four point interaction
changes if we include stubs in the three string vertex.
Both the original interaction and the one for the case of stubs
must have the same sign, because they only differ by the region of
integration over moduli space, and the integrand, as we
have seen, has a definite sign.
On the other hand, the interaction arising from the stubbed theory
would equal the original interaction plus a collection of Feynman graphs
with two three-string vertices and with one propagator whose proper
time is only partially integrated. Such terms, for completely integrated
propagators and massive intermediate fields would give a contribution
leading to a potential unbounded below. For partially integrated
propagators they also contribute  such kind of terms, both if the
field is truly massive, or if it is tachyonic. This indicates that
one should have expected an unbounded below elementary interaction.

It is now simple to explain why the choice of restricted
polyhedra (polyhedra with all nontrivial closed paths longer
than or equal to $2\pi$ [\kugokunitomo])  for closed
string vertices minimizes recursively the expansion coefficients of the
tachyon potential. We have seen that $\v\\3$ is minimized by
the Witten vertex. At the four point level we then have a missing region
$\V_{0,4}$. In the parametrization given by the final form in \trjk\ the
region of integration corresponding to $\V_{0,4}$ is fixed. At each
point in this region, the integrand, up to a constant, is given by
$1/\prod_i \rho_i^2 =  \exp (-4\pi \F)$, and as explained around Eqn.\nfu,
this quantity is minimized by the choice of coordinate disks determined
by the Strebel differential. Since the integrand is positive definite
throughout the region of integration, and, at every point is minimized
by the use of the Strebel differential, it follows that the integral is
minimized by the choice of the Strebel differential for the string vertex.
That is precisely the choice that defines the restricted polyhedron
corresponding to the standard four closed string vertex. It is clear that
the above considerations hold for any $\V_{0,N}$. The minimum value for
$\v\\N$ is obtained by using polyhedra throughout the region of integration.
Therefore, starting with the three string vertex, we are led recursively
by the minimization procedure to the restricted polyhedra of closed
string field theory.

\chapter{Estimates for the Tachyon Potential}

The present section is devoted to estimates of the tachyon potential.
While more analytic work on the evaluation of the tachyon potential
may be desirable, here we will get some intuitive feeling for the
growth of the coefficients $\v\\N$ for large $N$. The aim is to
find if the tachyon potential has a radius of convergence. We will
not be able to decide on this point, but we will obtain a series of
results that go in this direction.

For every number of punctures $N$, there is a
configuration of these punctures on the sphere for which
we can evaluate exactly
the measure of integration. This is the configuration where the punctures
are  ``equally separated" in a planar arrangement. These configurations
appear as a finite number of points in the boundary of $\V_{0,N}$, and
in some sense are the most problematic. The large $N$ behavior of the
measure at those points is such that if the whole integrand were to be
dominated by these points the tachyon potential would seem to have no
radius of convergence. The shape of $\V_{0,N}$ around those points, however, is
such that the contributions might be suppressed.

In each $\V_{0,N}$ there are configurations where the punctures are
distributed most symmetrically. It is intuitively clear that at these
configurations the measure is in some sense lowest. It is possible to
estimate this measure for large $N$, and conclude that, if dominated
by this contribution, the tachyon potential should have some radius
of convergence.

The behavior of the measure for the tachyon potential is such that
the measure grows as we approach degeneration, and if $\V_{0,4}$, for example,
was to extend over all of $\overline\M_{0,4}$ the naive integral would be
infinite. This infinity is not physical, because we do not expect
infinite amplitude for the scattering of four zero-momentum tachyons.
We explain how analytic continuation of the contribution from the
Feynman graphs removes this apparent contradiction.

We begin by presenting several exact results pertaining two, three, and
four-punctured spheres.

\section{Evaluation of Invariants}

Consider first the invariant $\chi_{12}$ of a sphere with two punctured
disks (Eqn.\twopmi). The disks may touch but they are assumed not to overlap.
Since the mapping radii can be as small as desired, the invariant
$\chi_{12}$ is not bounded above. It is actually bounded below, by
the value attained when we have a Strebel quadratic differential.
We can take the sphere punctured at zero and infinity, and the quadratic
differential to be $\varphi= -(dz)^2/z^2$. While this differential
does not determine a critical trajectory, we can take it to be
any closed horizontal trajectory, say $|z|=1$. This divides the $z$-sphere
into two disks, one punctured at $z=0$ with unit disk $|z|\leq 1$ and
the other punctured at $z= \infty$, or $w=0$, with $w=1/z$, and with
unit disk $|w| \leq 1$. It follows that their mapping radii are both
equal to one. We therefore have, using \tiinve\
$$\widehat\chi_{12}\equiv {\rm Min}\, (\chi_{12}) = {\rm Min} \Bigl( {1\over
\rho_{{}_{D_1(0)}}\,\rho_{{}_{D_2(\infty)}}}\Bigr) = 1\,.\eqn\tiinve$$

We now consider three punctured spheres and try to evaluate the minimum
value of the invariant $\chi_{123}$ defined in Eqn.\threp.
This minimum is achieved for the
three punctured sphere corresponding to the Witten vertex. In this vertex
we can represent the sphere by the $z$ plane, with the punctures at
$e^{\pm i\pi/3}, -1$. The disk surrounding the puncture at $z=e^{i \pi /3}$
is the wedge domain $0\leq {\rm Arg}(z) \leq 2\pi/3$. This domain is mapped to
a unit disk $|w|\leq 1$ by the transformations
$$t = z^{3/2} \,,\quad w = {t-i\over t+1}\, .\eqn\mapsnes$$
One readily finds that the mapping radius $\rho$ of the disk is given
by $\rho = |{dz\over dw}|_{w=0} = 4/3$. Furthermore, the distance
$|z\\I-z\\J|$ between any of the punctures is equal to $\sqrt{3}$. Therefore
back in \threp\ we obtain
$$\widehat\chi_{123} \equiv
{\rm Min}\, (\chi_{123})= \Bigl( {3\sqrt{3}\over 4} \Bigr)^3\, . $$
This result implies that the minimum possible value for
$|\v\\3|$ is realized with $\v\\3 = -3^9/2^{11}$ (see Eqn.\tprjk).

Another computation that is of interest is that of the most symmetric
four punctured sphere, a sphere where the punctures are at $z=0,1,\infty$,
and at $z =\rho= e^{i\pi/3}$. The Strebel quadratic differential for this
sphere can be found to be
$$\varphi =   { (2z+ 1)(2z- \rho) (2z-\bar\rho )
\over (z-1)^2 (z-\rho^2)^2(z-\bar\rho^2)^2 }\cdot {(dz)^2
\over z^2 }\, .\eqn\qdmss$$
Here the poles are located at the points $z=0, \,1,\, \rho^2$, and
$\bar\rho^2$.
The zeroes
are located at $z= -\half,\,\half \rho, \,\half\bar\rho$, and $\infty$.
One can use this expression
for a calculation of the mapping radii.

\section{The Measure at the Planar Configuration}

In each $\V_{0,N}$, for $N\geq 4$ there is a set of symmetric
planar configurations
for the punctures. They correspond to the surfaces obtained by  Feynman
diagrams constructed using only the three string vertex, and with all
the propagators collapsed with zero twist angle. We will consider the
case of $N$ punctures and give an exact evaluation for the measure.
This will be done in the frame where three punctures are mapped to the
standard points $z=0,1,\infty$, and the rest of the punctures will
be mapped to the points $z_1,z_2,\cdots z\\{N-3}$ lying on the real line
in between $z=0$ and $z=1$. Consequently, we will take $z\\{N-2}=1,
z\\{N-1}=\infty$ and $z\\N=0$. The measure we will calculate will be defined as
$$d\mu\\N \equiv \prod_{I=1}^{N-3}
{\d x\\I \d y\\I\over \rho\\I^2}\,\,
{|z\\{N-2}-z\\{N-1}|^2|z\\{N-2}-z\\N|^2 |z\\{N-1}-z\\N|^2
\over \rho\\{N-2}^2 \rho\\{N-1}^2 \rho\\N^2}\, ,\eqn\demeaw$$
which, up to constants, is the measure that appears in \wknoofffx.
The result will be of the form
$d\mu\\N = f\\N \prod_{k=1}^{N-3} \d x_k \d y_k$,
where $f\\N$ is a number depending on the number of punctures.

We begin the computation by using a $\xi$ plane where we place all the $N$
punctures equally spaced on the unit circle $|\xi|=1$. We thus let $\xi_k$
be the position of the $k$-th puncture, with
$$\xi_k = \exp \bigl( (2k-1) i\pi/N)\,,\quad k=1,2,\cdots N\,.\eqn\listp$$

\if y\figcount
\midinsert
\centerline{\epsffile{xiplane.eps}}
\nobreak
\narrower
\singlespace
\noindent
Figure 1. Planar configuration of punctures on the sphere. Shown are
the maps from the ring domain associated with a specific puncture to
the unit disk.
\medskip
\endinsert
\else\fi

In this presentation the ring domain surrounding a
puncture, say the first one, is the wedge domain $0\leq {\rm Arg}
(\xi) \leq 2\pi/N$ (see Figure 1) The mapping radius can be computed
exactly by mapping the wedge to the unit disk $|w| \leq 1$, via $t=
\xi^{N/2}$ and $w = {t-i\over t+1}$.  The result is $\rho= 4/N$, and
picking the three special punctures to be $\xi\\{N-2},\xi\\{N-1}$ and
$\xi\\N$, we find
$$\d\mu\\N = 64 \cdot \sin^4 \Bigl({\pi\over N}\Bigr)\cdot
 \sin^2 \Bigl({4\pi\over N}\Bigr)\cdot \Bigl({N\over 4}\Bigr)^{2N}\cdot
\prod_{k=1}^{N-3} \d^2\xi_k \,, \eqn\measup$$
where $\d^2\xi_k = \d\hbox{Re}\, \xi_k \,\d\hbox{Im} \,\xi_k $.
This is the measure, but in the $\xi$ plane. In order to transform it to
the $z$-plane we need the $\psl$ transformation that will satisfy
$z(\xi\\N) =0, z(\xi\\{N-1}) = 1$, and $z(\xi\\{N-2}) = \infty$. The desired
transformation is
$$ z = {\xi - e^{-i\pi/N} \over \xi - e^{-3i\pi/N}}\cdot \beta\,,\quad
{\rm with} \quad
|\beta| = {\sin (\pi/N)\over \sin (2\pi/N) }\,,  \eqn\heudk$$
and a small calculation gives
$$\d^2\xi_k  = \Bigl|{dz\over d\xi}\Bigr|^{-2}_{\xi_k}
\,\d x_k \d y_k= 4\cdot { \sin^2(2\pi/N)\over \sin^4 (\pi/N)}\cdot
\sin^4 \Bigl( {(k+1)\pi\over N}\Bigr)\cdot \d x_k \d y_k\,. \eqn\amst$$
This expression, used in \measup\ gives us the desired expression for
the measure
$$\eqalign{
\d\mu\\N &= \,\,4^{2N-3}\cdot \sin^4 \Bigl({\pi\over N}\Bigr)\cdot
 \sin^2 \Bigl({4\pi\over N}\Bigr)\cdot \Bigl({N\over 4}\Bigr)^{2N}\cdot
\Bigl({ \cos(\pi/N)\over \sin (\pi/N)}\Bigr)^{2N-6}\,\cr
&\quad\cdot\Bigl[\,\, \prod_{k=1}^{N-3}\sin^4
\Bigl( {(k+1)\pi\over N}\Bigr) \,\Bigr]
\cdot\prod_{k=1}^{N-3} \d x_k \,\d y_k \,.\cr}\eqn\finmeas$$
This is an exact result, valid for all $N\geq 4$. For the case of
$N=4$ it gives $\d\mu\\4 = 256 \d x \d y$.\foot{For $N=4$ the measure
can also be calculated exactly for the configuration with cross ratio
equal to $(1+i\sqrt{3})/2$. One finds [\belopolsky] $\d\mu\\4 = {2^{11}\over
3^4\root 3 \of 2 }\,\d x\,\d y \approx 20.07 \,\d x\,\d y$. This correponds
to the measure at the ``center'' of $\V_{0,4}$ and is indeed much smaller
than the measure $256\,\d x\,\d y$ at the corners of $\V_{0,4}$.}
Let us now consider
the leading behavior of this measure as $N\to \infty$.
The only term that requires some calculation is the product
$a\\N \equiv\prod_{k=1}^{N-3}\sin^4
( {(k+1)\pi/ N})$. One readily finds that as $N\to \infty$
$$\ln a\\N \,\sim\, 4\cdot {N\over \pi}
\int_0^\pi d\theta \ln (\sin \theta) = -4N \ln (2) \,\quad
\Rightarrow \quad a\\N \sim 4^{-2N} ,\eqn\estimp$$
and using this result, we find the large $N$ behaviour of the planar
measure
$$\d\mu\\N \,\sim\, 4^5\cdot\pi^6\cdot
 \Bigl[ {N^2\over 4\pi} \Bigr]^{2N-6}
\cdot\prod_{k=1}^{N-3} \d x_k \d y_k \,. \eqn\lnb$$
This was the result we were after. We see that this measure grows
like $N^{4N}$. This growth is so fast that presents an obstruction
to a simple proof of convergence for the series defining the tachyon
potential. Indeed, a very naive estimation would not yield convergence.
Let us see this next.

Let us assume that this planar uniform configuration is
indeed the point in $\V_{0,N}$  for which the
measure is the largest. This statement requires explanation, since
the numerical coefficient appearing in front of a measure can be
changed by $\psl$ transformations. Thus given any other configuration
in $\V_{0,N}$
with a puncture at $0,1$ and $\infty$ we do a transformation $z\to az$  with
$a= 1/z_{\rm max}$, where $z_{\rm max}$ is the position of the puncture
farther away from the origin. In this way we obtain a configuration with
all the punctures in the unit disk, the same two punctures at zero and
infinity, and some puncture at one. At this point the measures can
be compared and we expect the planar one to be larger. It is now clear
from the construction that the full $\M_{0,N}$ is overcounted if we fix
two punctures, one at zero, the other at infinity, and among the rest we
pick one at a time to be put at one, while the others are integrated all
over the inside of the unit disk. If we estimate this integral
using our value for the measure in the worst configuration we get
that each coefficient $\v\\N < [(N-2)\pi^{N-3}] N^{4N-12}$, where
the prefactor in brackets arises from the above described integrals
(this prefactor
does not really affect the issue of convergence).
The  growth $N^{4N}$ rules out
the possibility of convergence. This
bound is quite naive, but raises the possibility that there
may be no  radius of convergence for the tachyon potential.

\section{The Most Uniform Distribution of Punctures}

The corners of $\V_{0,N}$ turned out to be problematic. Since we expect
the measure for the coefficients of the tachyon potential to be lowest
at the most symmetric surfaces, we now estimate the measure at this point
in $\V_{0,N}$ for large $N$. The estimates we find are
consistent with some radius of convergence for the tachyon potential if
the integrals are dominated by these configurations.

It is possible to do a very simple estimate. To this end consider
the $z$ plane and place one
puncture at infinity with $|z|\geq 1$ its unit disk. In this way its
mapping radius is just one. All other punctures will be distributed
uniformly inside the disk. Because of area constraint we can imagine
that each puncture will then carry a little disk of radius $r$, with
$N\pi r^2 \sim \pi$ fixing the radius to be $r\sim 1/\sqrt{N}$.
The mapping
radius of each of these disks will be $r$. Another of the disks
will be fixed at $0$, and another to $2r$. We can now
estimate the measure, which is the integrand in \wknoofffx, where in
dealing with the three special punctures we make use of \threpp. We have then
$$ d\mu_{\rm sym} \sim {(2r)^2
\over r^2\cdot r^2 \cdot 1}\prod_{I=1}^{N-3}
{\d x\\I \d y\\I\over r^2}\,\sim N^{N-2}\prod_{I=1}^{N-3}
\d x\\I \d y\\I\,.
\eqn\wmeffx$$
Since all the punctures, except for the one at infinity, are
inside the unit disk, we can compare the measure given above
with the measure in the planar
configuration. In that case the measure coefficient went like $N^{4N}$ and
now it essentially goes like $N^N$, which is much smaller, as we
expected.\foot{Notice that if the punctures in the planar configuration
had remained in the boundary of the unit disk then the measure would
have only diverged like $N^{2N}$. The conformal map that brought them
all to the real line between $0$ and $1$ introduced an extra factor of
$N^{2N}$. This suggests that the divergence may actually not be as
strong as it seems at first sight.} We can also repeat the
estimate we did for the integration
over moduli space for the planar configuration, and again, we just get an extra
multiplicative factor of $N$, which is irrelevant. Therefore, if we
assume this configuration dominates we find $\v\\N\sim N^N$ and
the tachyon potential $\sum {\v\\N\over N!}\tau^N$ would have some radius
of convergence.

\section{Analytic Continuation and Divergences}

Here we want to discuss what happens if we do string field theory
with stubs of length $l$. It is well-known that as the length  of
the stubs goes to infinity $l\to \infty$ then the region of moduli
space corresponding to $\V_{0,4}$ approaches the full $\M_{0,4}$.
In this case the coefficient of the quartic
term in the potential will go into the full off-shell amplitude for
scattering of four tachyons at zero momentum.
We will examine the measure of integration for the tachyon potential
as we approach the boundaries of moduli
space and see that we would get a divergence corresponding to a tachyon
of zero momentum propagating over long times. We believe this divergence
is unphysical, and that the correct approach is to define the amplitude
by analytic continuation from a region in the parameter space of the
external momenta where the amplitude converges. When the full off-shell
amplitude is built from the vertex contribution and the Feynman diagram
contribution, analytic continuation is necessary for the Feynman part.

We therefore examine the off-shell formula for the evaluation of the
four string vertex for general off-shell tachyons. What we need is the
expression given in Eqn.\fxyu\ integrated over $\V_{0,4}$
$$A_{p_1\cdots p_4}  = {2\over \pi}\int_{\V_{0,4}}
\d x\,\d y \,{ |z|^{2p_1p_2} |z-1|^{2p_1p_3} \over \rho_z^{2-p_1^2}
\rho_0^{2-p_2^2}\rho_1^{2-p_3^2}\rho_\infty^{2-p_4^2}}\,, \eqn\fxyuu$$
where we have added subscripts to the mapping radii in order to
indicate the position of the punctures. Let us now examine what happens as
we attempt to integrate with $z\to 0$, corresponding to a degeneration where
punctures one and two collide. In this region $\rho_1$ and $\rho_\infty$
behave as constants, and we have that
$$A_{p_1\cdots p_4}  \sim \int_{|z|< c}
\d x \,\d y \,{ |z|^{2p_1p_2}  \over \rho_z^{2-p_1^2}
\rho_0^{2-p_2^2}}\,, \eqn\fxyuv$$
As the puncture at $z$ is getting close to the puncture at zero it is
intuitively clear that the mapping radii $\rho_z\sim \rho_0 \sim |z|/2$
as these are the radii of the ``largest'' nonintersecting disks surrounding
the punctures. Therefore
$$A_{p_1\cdots p_4}  \sim \int_{|z|< c}
 \,{ \d x \,\d y  \over  |z|^{4-p_1^2- p_2^2-2p_1p_2} }\,
=  \int_{|z|< c}
\,{  \d x \,\d y  \over  |z|^{4-(p_1+ p_2)^2}} \,,\eqn\fxyyuv$$
and we notice that the divergence is indeed controlled by the momentum
in the intermediate channel. If all the momenta were set to zero before
integration, we get a divergence
of the form $\int \d r/r^3$. But the way to proceed is to do the integral
in a momentum space region where we have no divergence
$$A_{p_1\cdots p_4}  \sim  \int_{|z|< c}
\,{  \d r \over  r^{3-(p_1+ p_2)^2}} \sim {1\over [2-(p_1+ p_2)^2] }
\,,\eqn\txyyuv$$
and the final result does not show a divergence for $p_1=p_2=0$.
Notice also that the denominator in the result is nothing else than
$L_0+\bar L_0$ for the intermediate tachyon, if that tachyon were
on shell, we would get a divergence due to it.

\chapter{A Formula for General Off-Shell Amplitudes on the Sphere}

In this section we will derive a general formula for off-shell
amplitudes on the sphere. In string field theory those amplitudes
are defined as  integrals over  subspaces $\A_N$ of the moduli space
$\hP_{0,N}$ of $N$-punctured spheres with local coordinates, up to phases,
at the punctures. We will assume a real parameterization of $\A_N$
and derive an operator expression which
expresses the amplitude as a multiple integral over these parameters.
The new point here is the explicit description of the
relevant antighost insertions necessary to obtain the integrand, and
the discussion explaining why the result does not depend neither on the
parameterization of
the subspace nor on the choice of a  global
uniformizer on the sphere.

We also give a formula for the case when the space $\A_N$ is parametrized
by complex coordinates.  For this case we will emphasize the
analogy between the $\b$--insertions for moduli and the $\b$--insertions
necessary to have $\psl$ invariance.
Finally, we will show how the general formula works by re-deriving the
off--shell Koba-Nielsen amplitude considered earlier.

\section{An Operator Formula for N--String Forms on the Sphere}

Recall that the state space $\H$
of closed string theory consists of the states
in the conformal theory that are annihilated both by $L_0-\bar L_0$ and by
$b_0-\bar b_0$.
Following [\zwiebachlong] we now
assign an $N$--linear function on $\H$
to any subspace $\A$ of $\wh\P_{0,N}$. The multilinear function is defined
as an integral over $\A$ of a canonical differential form
$$\{\Psi_1,\Psi_2,\dots,\Psi\\N\}_\A=
\int_\A{\Omega_{}^{}}^{(\dim\A-\dim\M_{0,N})0,N}_{\Psi_1\Psi_2\dots\Psi\\N}\,.
\eqn\general$$
One constructs the forms by verifying that suitable forms
in $\P_{0,N}$ do lead to well defined forms in $\wh\P_{0,N}$. This is
the origin of the restriction of the CFT state space to $\H$.
The canonical $2(N-3)+k$--form $\Omega^{(k)0,N}$ on $\wh\P_{0,N}$
is defined by its
action on $2(N-3)+k$ tangent vectors $V\\I\in T_{\Sigma_P}\wh\P_{0,N}$ as
$$\Omega^{(k)0,N}_{\Psi_1,\Psi_2,\dots\Psi\\N}(V_1,\dots,V_{2(N-3)+k})=
\fact
\bra{\Sigma_P}\b(\sv_1)\cdots\b(\sv_{2(N-3)+k})\ket{\Psi^N}\,.\eqn\bins$$
Here the surface state $\bra{\Sigma_P}$ is a bra living in
$(\HH^*)^{\otimes N}$ and represents the punctured Riemann surface $\Sigma_P$.
The symbol $\sv\\I$ denotes a Schiffer variation representing the tangent
$V\\I$, and
$${\bf b}({\bf v})=\sum_{I=1}^N
\oint_{w\\I=0}{{\rm d} w\\I\over 2\pi i}v^{(I)}(w\\I)b^{(I)}(w\\I)+
\oint_{\bar w\\I=0}{{\rm d} \bar w\\I\over 2\pi i}
\bar v^{(I)}(\bar w\\I)\bar b^{(I)}(\bar w\\I)\,.\eqn\bfb$$
Recall that a Schiffer variation for an $N$-punctured surface (in our present
case a sphere) is an $N$--tuple of vector fields
$\sv=\left(v^{(1)},v^{(2)},\cdots,v^{(N)}\right)$
where the vector $v^{(k)}$ is a  vector field defined in
the coordinate patch around the $k$--th puncture.\foot{In
general the vector fields $v^{(k)}$ are defined on some annuli
around the punctures and do not extend holomorphically to the whole
coordinate disk, in order to represent the change of modulus of the
underlying non-punctured surface (see [\zwiebachlong]). For $g=0$
the underlying surface is the Riemann sphere and has no moduli. Therefore, the
Schiffer vectors can be chosen to be extend throughout the coordinate disk.}
Let $w_k$ be the local coordinate around the $k$--th puncture $P_k$
($w_k(P_k)=0$). The variation  defines a new $N$--punctured Riemann sphere
with a new  chosen local coordinates $w_k' = w_k+\e v^{(k)}(w_k)$.
The new $k$-th puncture is defined to be at the point $P_k'$ such that
$w_k'(P_k') =0$. It follows that the $k$--th puncture is shifted
by $-\e v^{(k)}(P_k)$. For any tangent $V\in T_P\wh\P_{0,N}$
there is a corresponding Schiffer vector. Schiffer vectors are unique
up to the addition of  vectors that arise from the restriction of
holomorphic vectors on the surface minus the punctures.

Note that the insertions $\b(\sv)$ are invariantly defined, they only
depend on the Schiffer vector, and do not depend on the local
coordinates. Indeed $b$ is a primary field of conformal dimension $2$
or a holomorphic $2$--tensor. Being multiplied by a  holomorphic
vector field $v$ it produces a
holomorphic $1$--form, whose integral
$\oint_{w=0}{{\rm d} w\over 2\pi i}v(w)b(w)$
is well-defined and independent of the contour of integration.

In order to evaluate \general\ we choose some real
coordinates $\l_1,\dots,\l_{\dim\A}$ on $\A$.  Let $\{V_{\l_k}\}=
\partial/\partial \l_k$ be the corresponding tangent
vectors, and let $\{\d\l_k\}$ be the corresponding dual one-forms,
 \ie\ $\d\l_k(V_{\l_l})=\delta_{k,l}$.
Using $\{V_{\l_k}\}$ we can rewrite \general\ as
$$\{\Psi_1,\cdots,\Psi\\N\}_\A=\fact
\int\d\l_1\cdots\d\l_{{\dim\A}}
\bra{\Sigma_P}\b(\sv_{\l_1})\cdots\b(\sv_{\l_{\dim\A}})
\ket{\Psi_1}\cdots\ket{\Psi\\N}\,.\eqn\lint$$
In order to continue we must parameterize $\A$ as it sits in the moduli space
$\wh\P_{0,N}$. Let $w\\I$ be a local coordinate around the $I$--th puncture.
Given  a global uniformizer $z$ on the Riemann sphere we can represent
$w\\I$ by as an invertible analytic function $w\\I(z)$ defined on some disk
in the $z$--plane which maps the disk to a standard unit disk $|w\\I|<1$. The
inverse map $h\\I=w\\I^{-1}$ is therefore an analytic function on a unit
disk. Therefore, $N$ functions $h\\I(w\\I)$ define a point in $\wh\P_{0,N}$,
the sphere with $N$ punctures at $h\\I(0)$ and local coordinates given by
$w\\I(z)=h\\I^{-1}(z)$.  The embedding of $\A$ in $\wh\P_{0,N}$
is then represented by a set of $N$ holomorphic functions
parameterized by the real coordinates $\l_k$ on $\A$:
$\{h_1(\{\l_k\};w_1),\cdots,h\\N(\{\l_k\};w\\N)\}$.
It is well known how to write the  state $\bra{\Sigma_P}$ in terms of
$h\\I$'s (see [\leclair]).
$$\eqalign{
\bra{\Sigma_P}=\,\, &2\cdot\int\prod_{I=1}^N \d p\\I\,
      (2\pi)^D\delta^D\left(\sum p\\I\right)
      \bigotimes_{I=1}^N\bra{{\bf 1}^c,p\\I}
      \int\d^2\zeta^1\d^2\zeta^2\d^2\zeta^3\cr
      &\cdot\ds\exp\Big(E(\alpha)+F(b,c)
         -\sum_{n=1}^3\sum_{m\geq -1}\left(
              \zeta^n M^{nJ}_{m}b^{(J)}_m -
      \bar\zeta^n\ov{{M}^{nJ}_{m}}\bar b^{(J)}_m
      \right)\Big).}\eqn\sigmap$$
where repeated uppercase indices $I, J \cdots $, are summed over the $N$
values they take. In here
$$\eqalign{
         \ds E(\alpha)&=-{1\over2}\sum_{n,m\geq0}\left(
                   \alpha^{(I)}_n N^{IJ}_{nm}\alpha^{(J)}_m+
                   \bar\alpha^{(I)}_n \ov{N^{IJ}_{nm}}\bar\alpha^{(J)}_m
               \right)\,,\cr
         \ds F(b,c)&=\sum_{n\geq2\atop m\geq -1}\left(
              c^{(I)}_n\tilde N^{IJ}_{nm}b^{(J)}_m+
              \bar c^{(I)}_n\ov{\tilde{N}^{IJ}_{nm}}\bar b^{(J)}_m
           \right)\,.\cr}\eqn\efdef$$
The vacua satisfy $\braket{{\bf 1}^c,p}{{\bf 1},q}=\delta(p-q)$, and
the odd Grassmann variables $\zeta^n$ are integrated using
$$\int\d^2\zeta^1\d^2\zeta^2\d^2\zeta^3~\,\zeta^1\bar\zeta^1
            ~\zeta^2\bar\zeta^2 ~\zeta^3\bar\zeta^3 \equiv 1\,.\eqn\grint$$
Note that the effect of this integration is to give the product of
six antighost insertions coming from the last term in the exponential in
Eqn.\sigmap. The minus sign in front of this term is actually irrelevant
(as the reader can check) but it was included for later convenience.
A bar over a number means complex conjugate while a bar
over an operator is used in order to distinguish the left--moving
modes from right--moving ones.
The {\em Neumann coefficients} $ N_{mn}^{IJ}$ and
$\tilde{N}^{IJ}_{mn}$ are given by the following formulae:
$$
\eqalign{
    N_{00}^{IJ}&=\ds\left\{ {\log(h'\\I(0)),\hfill I=J\atop
      \log(h\\I(0)-h\\J(0)),\quad I\neq J}\right.\cr
    N_{0n}^{IJ}&=\ds{1\over n}\oint_{w=0} {\d w\over 2\pi i} w^{-n}
    h'\\J(w){-1\over h\\I(0)-h\\J(w)}\,,\cr
    N_{mn}^{IJ}&=\ds{1\over m}\oint_{z=0} {\d z\over2\pi i}z^{-m} h'\\I(z)
    {1\over n}\oint_{w=0}{\d w\over2\pi i} w^{-n}h'\\J(w)
    {1\over (h\\I(z)-h\\J(w))^2}\,,\cr
    \tilde{N}_{mn}^{IJ}&=\ds{1\over m}\oint_{z=0}{\d z\over2\pi i} z^{-m+1}
    (h'\\I(z))^2
    {1\over n}\oint_{w=0}{\d w\over2\pi i} w^{-n-2} (h'\\J(w))^{-1}
    {1\over h\\I(z)-h\\J(w)}\,.\cr}
\eqn\neumann$$
Moreover,
$$M_{m}^{nJ}  =\ds\oint_{w=0}{\d w\over 2\pi i} w^{-m-2} (h'\\J(w))^{-1}
    [h\\J(w)]^{n-1},\quad n=1,2,3\,.  \eqn\mneu$$

\noindent
\underbar{Antighost Insertions.}
Now let us show how to take the $\b$ insertions
into account. In order to calculate the $\b$ insertion associated to a tangent
vector $V \in T\\\Sigma\wh\P_{0,N}$, we must find
the Schiffer vector (field) that realizes the deformation of the surface
$\Sigma$ specified by $V$.
Consider a line $c(t)$ in $\wh\P_{0,N}$ parameterized by the real
parameter $t$: $c:[0,1]\to\wh\P_{0,N} $, such that $\Sigma = c(0)$, and
the tangent vector to the curve
is $V=c_*\left({\d\over\d t}\right)$. We will see now how one can use this
setup to define in a natural way a vector field on the neighborhood of
the punctures of the Riemann surface $\Sigma$. This vector field is
the Schiffer vector.

\if y\figcount

\midinsert
\epsfxsize\hsize
\centerline{\epsffile{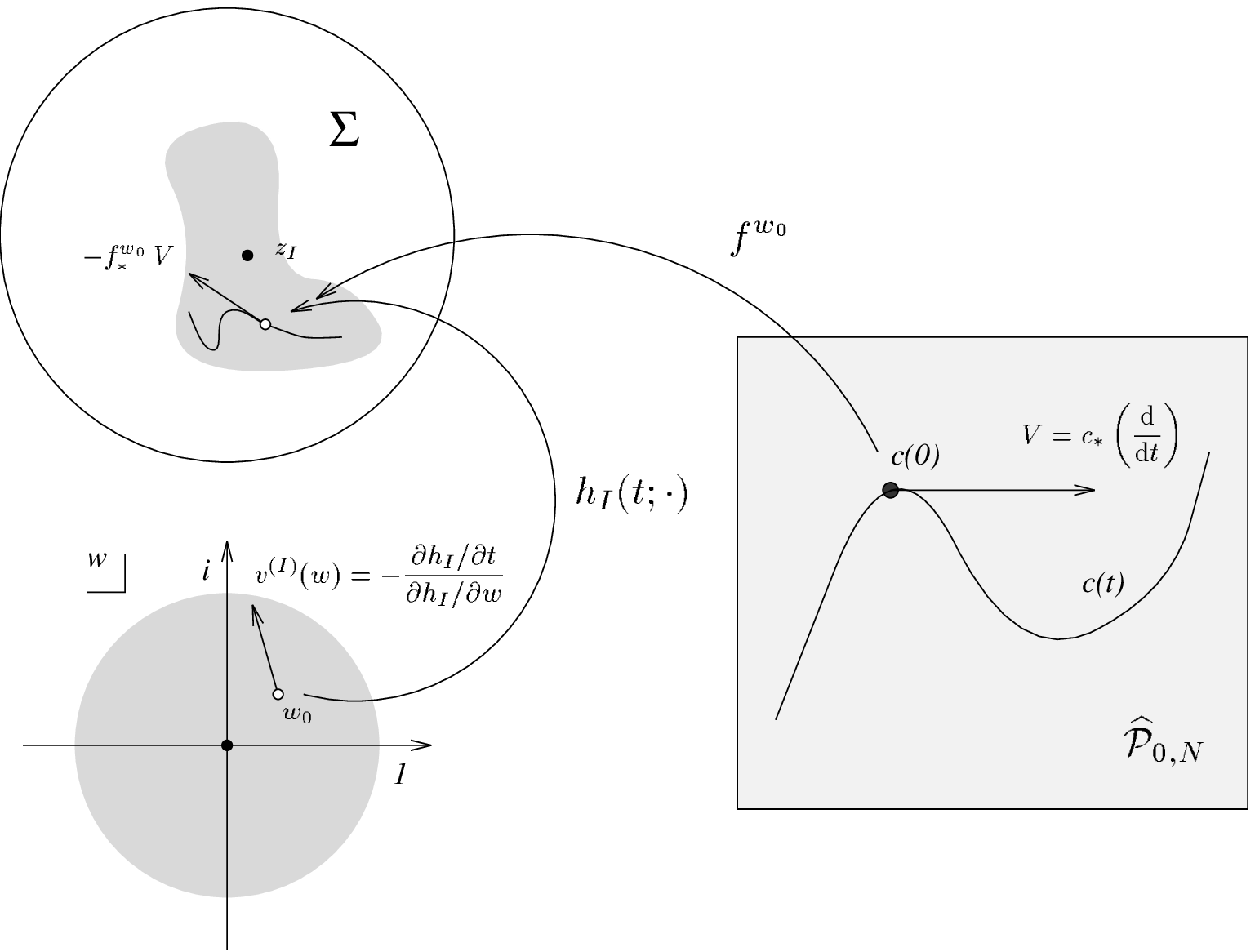}}
\nobreak
\narrower
\singlespace
\noindent
Figure 2. We show how to obtain a Schiffer vector field associated to
a tangent vector in $\wh\P_{0,N}$. Shown are the Riemann surface $\Sigma$
and the local coordinate plane $w$.
\medskip
\endinsert

\else\fi

We can represent the curve $c(t)$ by $N$
functions $h\\I(t;w)$, holomorphic in $w$, and parameterized by $t$.
Choose a fixed value $w_0$ of the $w$ disk.
We now define a map $f^{w_0}: c(t)\to \Sigma$ from the curve $c(t)$
to a curve on the surface $\Sigma$.
 The function $f^{w_0}$ takes
$c(t)$ to  $z\\I(t)= h\\I(t,w_0)$ for each value of $t$.
We can now use the map $f^{w_0}$ to produce a push-forward map of vectors
$f_*^{w_0}: Tc\to T_{h\\I(0,w_0)}\Sigma$. In this way we can produce the
vector $f_*^{w_0} V\in T_{z\\I(0,w_0)}\Sigma$. By varying the value of
$w_0$ we obtain a vector {\it field} on the neighborhood of the $I$-th
puncture. We claim that this vector field, with a minus sign,
is the Schiffer vector.
In components, and with an extra minus sign, the pullback gives
$$v^{(I)}(z)=-{\p h\\I\over\p t}\left(t;w\\I(z)\right)\,,\eqn\derschiffer$$
It is useful to refer the Schiffer vector to the local  coordinate $w\\I$.
We then find, by pushing the vector further
$$v^{(I)}(w\\I)=- \left({\p h\\I\over \p w\\I}\right)^{-1}
\cdot{\p h\\I(t;w\\I)\over \p t}.\eqn\tan$$
By definition, the Schiffer
vector $\sv(V)$ corresponding to the vector $V$ is given by the
collection of vector fields
$\sv=\left(v^{(1)}(w_1),\cdots,v^{(N)}(w\\N)\right)$.
If we define the  vector $V_{\l_k}\in T\wh\P_{0,N}$ to be the
tangent associated to the coordinate
curve parameterized by $\l_k$, we then write the Schiffer variation for
$V_{\l_k}$ as:
$$v_{\l_k}^{(I)}(w\\I)=-\,{1\over h\\I'(w\\I)}{\p h\\I(\l;w\\I)\over \p\l_k}\,,
\eqn\svk$$
where $h\\I'(w\\I) \equiv (\partial h\\I/\partial w\\I)$.

Before proceeding any further, let us confirm that the `natural' vector
we have obtained is indeed the Schiffer vector. This is easily done.
Let $p$ denote a point in the Riemann surface $\Sigma$ and let $w\\I(p)$
denote its local coordinate. By definition, the Schiffer vector defines
a new coordinate $w\\I'(p)$ as $w\\I'(p) = w(p) + dt \, v^I(w(p))$, where
$t$ is again a parameter for the deformation. Since the $z$-coordinate
of the point $p$ does not change under the deformation, we must have that
$h\\I (t+ dt , w\\I'(p)) = h\\I(t , w\\I(p))$. Upon expansion of this last
relation one immediately recovers Eqn.\tan.

One more comment is in  order. What happened to the usual ambiguity
in choosing Schiffer vectors?  Schiffer vectors are ambiguous since
there are nonvanishing $N$-tuples that do not induce any deformation.
This happens when the $N$-tuples can be used to define a holomorphic
vector on the surface minus the punctures. In our case the ambiguity is due
to the fact that the functions $h\\I(\l , w)$ can be composed with any
$\psl$ transformation $S$ in the form $S\circ h\\I(\l , w)$. We
will come back to this point later.

Using Eqn.\bfb\ we can now write $\b(\sv^\l)$ as
$$\b(\sv_{\l_k})=-\sum_{m\geq -1}
 ( B^{kJ}_m b^{(J)}_m+\ov{B^{kJ}_m}\bar b^{(J)}_m)\,,\eqn\tyu$$
where, as usual, the repeated index $J$ is summed over the number of punctures,
and
$$
B^{kJ}_{m}=\oint_{w=0}{\d w\over 2\pi i}\, w^{-m-2}
\,{1\over  h'\\J(w)}\, {\p h\\J(\l ,w)\over \p\lambda_k} \,.\eqn\MNeumann$$
The range $m\geq -1$ has been obtained because the Schiffer vectors
can be chosen to be holomorphic and not to have poles at the punctures
(this will not be the case for higher genus surfaces).

Let us now treat the $b$
insertions in a way similar to that used for the zero modes
in \sigmap. Let $\zeta\\I$ and $\eta\\I$ be anti-commuting
variables, then
$$\int
  \d\xi_1\dots\d\xi_n{\rm e}^{\xi_1\eta_1+\cdots+\xi_n\eta_n}=
  \int
  \d\xi_1{\rm e}^{\xi_1\eta_1}\cdots
  \int\d\xi_n{\rm e}^{\xi_n\eta_n}=
  \eta_1\cdots\eta_n.
    \eqn\trick$$
This observation allows us to represent the  product of $\b$ insertions in
\bins\ as an integral of an exponent.
$$\b(\sv_{\l_1})\cdots\b(\sv_{\l_{\dim\A}})=
\int\prod_{k=1}^{\dim\A}\d\xi^k\exp\Bigl(
-\sum_{k=1}^{\dim\A}\sum_{m\geq-1}\xi^k\left(B^{kJ}_m b^{(J)}_m+\ov{
B^{kJ}_m}\bar b^{(J)}_m
     \right)\,\Bigr)\,, \eqn\tricky$$
where  the $\xi^n$'s are real Grassmann odd variables. The multi-linear
product \general\ now assumes the form
$$
\eqalign{\ds
\{\Psi^N\}_\A&=2\left( {i\over 2\pi} \right)^{N-3}\int
      \prod_{I=1}^N \d^D p\\I
      (2\pi )^D\delta^D\left(\sum_{I=1}^Np\\I\right)
      \bigotimes_{I=1}^N\bra{{\bf 1}^c,p\\I}
      \int\prod_{k=1}^{\dim\A}\d\l_k
      \prod_{k=1}^{\dim\A}\d\xi^k\cr
      &\int\d^2\zeta^1\d^2\zeta^2\d^2\zeta^3
       \ds\exp\Bigg( E(\alpha) + F(b,c)
         \ds-\sum_{k=1}^3\sum_{m\geq -1}\left( \zeta^k M^{kJ}_{m}b^{(J)}_m
              -\bar\zeta^k\ov{M^{kJ}_{m}}\bar b^{(J)}_m \right)\cr
         & \qquad\qquad\qquad\qquad\qquad
\qquad\qquad\qquad\ds-\sum_{k=1}^{\dim\A}
           \sum_{m\geq -1}\xi^k\left(
              B^{kJ}_{m}b^{(J)}_m+
              \ov{B^{kJ}_{m}}\bar b^{(J)}_m
           \right)\Bigg)\ket{\Psi^N},}   \eqn\aterm$$
The above formula together with \efdef,\neumann, \mneu\ and \MNeumann\
gives a closed expression for a multi-linear form associated with
$\A\subset\wh\P_{0,N}$.

The resemblance of the last two terms appearing in the exponential
is not a coincidence. While the last term arose from the antighost
insertions for moduli, the first term, appearing already in the description
of the surface state $\bra{\Sigma}$, can be thought as the antighost
insertions due to the Schiffer vectors that represent $\psl$
transformations.  This is readily verified. Consider the sphere with
uniformizer $z$. The six globally defined vector fields are given by
$v_k(z) = z^k$, and $v_k'(z) = iz^k$ with $k=0,1,2$. Referring them
to the local coordinates one sees that $v_k^I(w) = [h\\I(w)]^k/h\\I'(w)$
and ${v'}_k^I(w) = i[h\\I(w)]^k/h\\I'(w)$. As a consequence
$${\bf b}(v_k) = \sum_{m\geq -1} \left( M^{kJ}_m\, b_m + \overline{M^{kJ}_m}
\,\bar b_m \right) \,,\eqn\sltwoc$$
$${\bf b}(v_k') = i\sum_{m\geq -1} \left( M^{kJ}_m\, b_m - \overline{M^{kJ}_m}
\,\bar b_m \right) \,,\eqn\slytwoc$$
where the $M$ coefficients were defined in Eqn.\mneu. It is clear that the
product of the six insertions precisely reproduces the effect of the first
sum in the exponential of Eqn.\aterm.

In order to be used in \aterm\ the subspace
$\A$ is parametrized by some
coordinates $\l_k$. The expression for the multilinear form is independent
of the choice of coordinates; it is a well-defined form on $\A$.\foot{This
is easily verified explicitly. Under coordinate transformations,
the product $\prod \d\lambda_i$ transforms with a Jacobian, and the product
of antighost insertions, as a consequence of \MNeumann\ transforms with
the inverse Jacobian.} Once the parametrization is chosen, the space $\A$
has to be described by the $N$ functions $h\\I(\lambda ,w)$.
These functions, as we move on $\A$ are defined up to a
{\it local} linear fractional
transformation. At every point in moduli space we are free to change
the uniformizer. Let us see why \aterm\ has local $\psl$ invariance.
The bosonic Neumann coefficients $N_{nm}^{IJ}$ are
$\psl$ invariant by themselves for $n,m\geq1$. The sums
$\sum_{IJ} \alpha^{(I)}_0 N^{IJ}_{00}\alpha^{(J)}_0=
\sum_{IJ}N^{IJ}_{00}p\\Ip\\J$ and $\sum_{J} N^{IJ}_{m0}\alpha^{(J)}_0=
\sum_{J}N^{IJ}_{m0}p\\J$ can be shown to be invariant due to
momentum conservation $\sum p\\J=0$. A detailed analysis of
$\psl$ properties of the ghost part of a surface state
have been presented by LeClair {\it et al.} in [\leclair]
for open strings. Their arguments can be readily generalized to our
case. The only truly new part that appears in \aterm\
is the last term in the exponential.

Under global $\psl$ transformations, namely transformations of the form
$h\to (ah+b)/(ch+d)$, with $a,b,c$ and $d$ independent of $\lambda$,
this term is invariant
because so is every coefficient $B^{kJ}_m$. Since a general local
$\psl$ tranformation can be written locally as a global transformation
plus an infinitesimal local one, we must now show invariance under
infinitesimal local transformations. These are transformations of the form
$$\widetilde h\\I = h\\I + a(\l) + b(\l)h\\I + c(\l) h\\I^2 \, ,\eqn\localtr$$
for $a,b$, and $c$ small.
A short calculation shows that
$$-{1\over \widetilde h\\J'}{\partial \widetilde h\\I\over \partial \lambda_k}
\,=\, -{1\over  h\\J'}{\partial h\\I\over \partial \lambda_k}\, - \,
{1\over  h\\J'} \Bigl[ {\partial a\over \partial \lambda_k} +
{\partial b\over \partial \lambda_k}h\\I +
{\partial c\over \partial \lambda_k}h\\I^2 \Bigr] \, . \eqn\trlocal$$
On the left hand side we have the new Schiffer vector, and the first
term the right hand side is the old Schiffer vector. We see that they
differ by a linear superposition of the Schiffer
vectors $v_k^I$ and ${v'}_k^I$ introduced earlier in our discussion of
$\psl$ transformations (immediately above Eqn.\sltwoc.) It follows that
the extra contributions they make to the antighost insertions vanish
when included in the multilinear form because the multilinear form already
includes the antighost insertions corresponding to the Schiffer vectors
generating $\psl$.

\noindent
\underbar{Complex Coordinates}
In some applications the subspace $\A$ has even dimension and can be
equipped with complex coordinates. Let $\dim^c\A$ denote the complex
dimension of $\A$
and let $\{\l_1,\cdots,\l_{\dim^c\A}\}$ be a set of complex coordinates.
The subspace $\A$ can now be represented by
the collection of functions
$\{h\\I(\{\l_k\},\{\bar\l_k\};w\\I)\}$ with $I=1,\cdots ,N$. In
order to derive a formula for this case we simply take the earlier
result for two real insertions and pass to complex coordinates. We
thus consider
$$ \Omega^2 = d\l_1\wedge d\l_2 \,\, d\xi^1 d\xi^2 \, \exp
\Bigl[\, -\sum_{k=1}^2 \xi^k\left(
              B^{kJ}_{m}b^{(J)}_m+
           \ov{B^{kJ}_{m}}\bar b^{(J)}_m\right) \Bigr]\, , \eqn\trans$$
Using complex coordinates $\l_1' = \l_1+ i\l_2$ and ${\xi'}^1 = \xi^1+ i
\xi^2$,
and letting $\int d^2 {\xi'}\, \xi' \bar \xi' \equiv 1$, we can write
the above as
$$ \Omega^2 = d\l'_1\wedge d\bar\l'_1 \,\, d^2 \xi'_1  \,
\exp \Bigl[\, -{\xi'}^1\left(
              B^{1 J}_{m}b^{(J)}_m+
           \ov{B^{\bar 1J}_{m}}\bar b^{(J)}_m\right)
 +\bar{\xi'}^1\left(
              B^{\bar 1J}_{m}b^{(J)}_m+
           \ov{B^{1J}_{m}}\bar b^{(J)}_m\right)\Bigr]\, , \eqn\tras$$
where we have defined
$$\eqalign{B^{ kJ}_{m}&=\ds
 \oint_{w=0}{\d w\over2\pi i}w^{-m-2}
      {1\over  h'\\J(w) } {{\p h\\J(\l,\bar\l;w)\over\p \l_k}},\cr
    B^{\bar k J}_{m}&=\ds
\oint_{w=0}{\d w\over2\pi i}w^{-m-2}
 {1\over  h'\\J(w) } {{\p h\\J(\l,\bar\l;w)\over\p\bar\l_k}}.\cr} \eqn\sdff$$
Note that the $B^{\bar k J}_{m}$ coefficients do not vanish because
the embedding of $\A$ in $\wh\P_{0,N}$ need
not be holomorphic and, as a consequence,  the derivatives
$\p h\\I/\p\bar\l_k$ need  not vanish. Using this
result we can now  rewrite \aterm\ for the case of complex coordinates
for moduli
$$\eqalign{\ds
\{\Psi^N\}_{\A}=2&\fact
  \hskip-6pt \int\prod_{I=1}^N\d^Dp\\I(2\pi)^D\delta^D\left(\sum p\\I\right)
  \bigotimes_{I=1}^N\bra{{\bf 1}^c,p\\I}
      \int_\A\prod_{k=1}^{\dim^c\A}\d\l_k\wedge\d\bar\l_k
      \prod_{k=1}^{\dim^c\A}\d^2\xi^k \cr
        &\prod_{k=1}^3 \hbox{d}^2\zeta^k \,\ds
      \exp\Bigg(E(\alpha)+F(b,c)
      -\ds\sum_{k=1}^3\Big(
                         \zeta^kM^{kJ}_{m}b^{(J)}_m  -
                         \bar\zeta^k\ov{M^{kJ}_{m}}\bar b^{(J)}_m\Big)\cr
   & \qquad -\ds\sum_{k=1}^{\dim^c\A}\Big[ \xi^k\Big(
                     B^{ kJ}_{m}b^{(J)}_m+
                     \overline{B^{\bar k J}_{m}}\bar b^{(J)}_m\Big)
    -\ds\bar\xi^k\Big(
                     B^{\bar kJ}_{m}b^{(J)}_m+
                     \overline{B^{ k J}_{m}}\bar b^{(J)}_m
      \Big)\Big] \Bigg)\ket{\Psi^N},}\eqn\atermc$$
where $m\geq -1$ for the implicit oscillator sum.
It is useful to bring out the similarity between the antighost
insertions for $\psl$ and those for moduli. In order to achieve
this goal we introduce
$\xi^{\dim^c\A+ k}\equiv \zeta^k$ for $k=1,2,3$.
$$\eqalign{\ds
\{\Psi^N\}_{\A}=2&\fact
  \hskip-6pt \int\prod_{I=1}^N\d^Dp\\I(2\pi)^D\delta^D\left(\sum p\\I\right)
  \bigotimes_{I=1}^N\bra{{\bf 1}^c,p\\I}
      \int_\A\prod_{k=1}^{\dim^c\A}\d\l_k\wedge\d\bar\l_k \cr
        &\prod_{k=1}^{\dim^c\A+ 3}\hskip-8pt\d^2\xi^k \cdot\ds
      \exp\Bigg(E(\alpha)+F(b,c)
      -\ds\hskip-8pt\sum_{k=1}^{\dim^c\A+ 3}\Big(
                         \xi^k\B^{kJ}_{m}b^{(J)}_m  -
                         \bar\xi^k\ov{\B^{kJ}_{m}}\bar b^{(J)}_m\Big)\cr
   & \qquad -\ds\sum_{k=1}^{\dim^c\A}\Big( \xi^k\,
                    \overline{B^{\bar k J}_{m}}\,\bar b^{(J)}_m
    -\ds\bar\xi^k
                     B^{\bar kJ}_{m}b^{(J)}_m
      \Big) \Bigg)\ket{\Psi^N},}\eqn\atermcc$$
where the script style $\B$ matrix elements are defined as
$$
\B^{kJ}_{m}=
\cases{B^{kJ}_m\, , & for $k\leq\dim^c\A$ \cr
       M^{(k-\dim^c\A)J}_m\, , &
                           for $k-\dim^c\A=1,2,3$. \cr}\eqn\ems$$
This concludes our construction of off-shell amplitudes as
forms on  moduli spaces of punctured spheres.

\section{Application to Off-Shell Tachyons}

Let us see how the formulae derived above work
for the case of
$N$ tachyons with arbitrary momenta. This particular example
allows us to confirm our earlier calculation of off-shell tachyon
amplitudes. Specifically, we are going to evaluate the multilinear function
$\{\tau_{p_1},\cdots,\tau_{p\\N}\}$
where $\ket{\tau_{p_i}}=c_{1}\bar{c}_{1}\ket{{\bf 1}, p_i}$. In this
case the state to be contracted with the bra representing the multilinear
function is
$\ket{\tau^N}=\otimes_{I=1}^Nc_1^{(I)}\bar c_1^{(I)}\ket{{\bf 1},p\\I}$.
Upon contraction with this state we will only get contributions
from $b^{(I)}_{-1}$, $\bar{b}^{(I)}_{-1}$, and
the  matter zero modes $\alpha_0^{(I)}=\bar\alpha_0^{(I)}=i p\\I$.

We will use as moduli the complex coordinates $z_1, \cdots , z\\{N-3}$
representing the position of the first $(N-3)$ punctures. Therefore,
$\l_k = z_k$, for $k=1, \ldots , N-3$, and we must use Eqn.\atermcc\
to calculate the multilinear function.  Our setting of the $z$-coordinates
as moduli implies that the functions $h\\J$ take the form
$$h\\J (z,\bar z , w) = z\\J  + a(z ,\bar z) w + \cdots \, .\eqn\basicform$$
It then follows that
$$M_{-1}^{kJ}  =\ds\oint_{w=0}{\d w\over 2\pi i}\, {1\over w} (h'\\J(w))^{-1}
    [h\\J(w)]^{k-1} \, = \,{z\\J^{k-1} \over {h'}\\J(0) }\, ,  \eqn\xneue$$
and furthermore
$$B^{kJ}_{-1}=\oint_{w=0}{\d w\over 2\pi i}\, {1\over w}
\,{1\over  h'\\J(w)}\, {\p h\\J\over \p\lambda_k}\,=\,
{\delta^{kJ}\over{h'}\\J(0) } \,,\eqn\xeumann$$
while $B^{\bar k J}_m = 0$, since $\p h\\J/\p \bar z_k = 0$.
With this information, back in \atermcc\ we find
$$\eqalign{\ds\{\tau_{p_1},\cdots,\tau_{p\\N}\}_{\A}=2&\fact
  \hskip-6pt (2\pi)^D\delta^D ({\bf 0} )
  \int\prod_{k=1}^{N-3}\d z_k\wedge\d\bar z_k\,\,\prod_{k=1}^N
\d^2\xi^k  \cr
        &\cdot\bra{{\bf 1}^c}
      \exp\Bigg(E(\alpha)
      -\ds\sum_{k=1}^N\Big(
                         \xi^k\B^{kJ}_{-1}b^{(J)}_{-1}  -
                         \bar\xi^k\ov{\B^{kJ}_{-1}}\bar b^{(J)}_{-1}\Big)
 \Bigg)\prod_{I=1}^N c_1^{(I)}\bar c_1^{(I)}\ket{{\bf 1}}\,. \cr}\eqn\atermtt$$
We can now calculate the bosonic contribution from $E(\alpha)$
$$\eqalign{
\exp (E(\alpha)) &=
\exp\Big(-{1\over2}\sum_{I,J=1}^N
\left(\alpha_0^{(I)}N_{00}^{IJ}\alpha_0^{(J)}+
      \bar\alpha_0^{(I)}\ov{N_{00}^{IJ}}\bar\alpha_0^{(J)}\right)\Big)\cr
& = \exp\Big(-{1\over2}\sum_{I,J=1}^N
      \left(N_{00}^{IJ}+\ov{N_{00}^{IJ}}\right) p\\I p\\J\Big)\cr
&=\prod_{I<J}\left(
              {|h\\I(0)-h\\J(0)|^2\over |h'\\I(0)|\cdot|h'\\J(0)|}
           \right)^{p\\Ip\\J} = \prod_{I<J} \, \chi\\{IJ}^{p\\Ip\\J}\,, \cr}
\eqn\evalalpha$$
where we have used the expression for Neumann coefficients \neumann,
momentum conservation, and \twopmi. We thus obtain
$$\eqalign{\ds\{\tau_{p_1},\cdots,\tau_{p\\N}\}_{\A}=\,
&{2\over \pi^{N-3}}\, (2\pi)^D\delta^D ({\bf 0} )
  \int\prod_{k=1}^{N-3}\d x_k\, \d y_k\,
\prod_{I<J} \, \chi\\{IJ}^{p\\Ip\\J}\,  \cr
        &\cdot\prod_{k=1}^{N}\d^2\xi^k \bra{{\bf 1}^c} \exp\Big(
      -\ds\sum_{k=1}^N\Big(
                         \xi^k\B^{kJ}_{-1}b^{(J)}_{-1}  -
                         \bar\xi^k\ov{\B^{kJ}_{-1}}\bar b^{(J)}_{-1}\Big)
 \Big)\prod_{I=1}^N c_1^{(I)}\bar c_1^{(I)}\ket{{\bf 1}}\,. \cr}\eqn\atermttt$$
Consider the second line of the above equation. The effect of the
ghost state is to select the term in the exponential proportional
to the product of all antighosts. Since
$$\bra{{\bf 1}^c}
\prod_{I=1}^N b_{-1}^{(I)}\bar b_{-1}^{(I)}
\prod_{I=1}^N c_1^{(I)}\bar c_1^{(I)}\ket{{\bf 1}}= (-)^N\, ,\eqn\auxil$$
we can write the second line of \atermtt\ using a second set of
Grasmmann variables $\eta^k$
$$(-)^N\int\prod_{k=1}^{N}\d^2\xi^k \,\d^2\eta^k
      \exp\Big[
      \ds\sum_{k,p=1}^N\Big(
                         -\xi^k\,\B^{kp}_{-1}\,\eta^p  +
                     \bar\xi^k\,\ov{\B^{kp}_{-1}}\,\bar \eta^p\Big)\Big]
\,=\,(-)^N\, \det (\B) \, \det (\bar \B )\,, \eqn\detprop$$
where in the last step we used a standard formula in Grassmann integration.
We can now use Eqns. \xneue, \xeumann, and \ems\ to calculate $|\det \B \,|^2$.
We find
$$\eqalign{
|\det \B |^2 &= |(z\\N -z\\{N-2}) (z\\N -z\\{N-1}) (z\\{N-2} -z\\{N-1})|^2
\prod_{I=1}^N {1\over |{h'}\\I (0)|^2}\, , \cr &
= \chi^2\\{N-2,N-1,N}\prod_{I=1}^{N-3} {1\over \rho\\I^2} \, , \cr}\eqn\ambv$$
where we made use of the definition of the mapping radius and of
Eqn.\threp. We can now assemble the final form of the tachyon
multilinear function. Back in \atermttt\ we have
$$\{\tau_{p_1},\cdots,\tau_{p\\N}\}_\A=(-)^N{2\over\pi^{N-3}}\int
                      \prod_{I=1}^{N-3}{\d x\\I \d y\\I \over\rho\\I^2}
                     \, \chi^2_{N-2,N-1,N}\prod_{I<J}\chi\\{IJ}^{q\\Iq\\J}\cdot
(2\pi)^D \delta ({\bf 0})\,. \eqn\finalf$$
which agrees precisely with the off-shell Koba-Nielsen formula \knofffx.

\ack We are grateful to G. Moore who stimulated our interest in the
computation of the tachyon potential, and gave us his result as
quoted in Eqn.\moore. We also wish to acknowledge instructive discussions
with  T. Banks, A. Sen, H. Sonoda and C. Thorn.

\APPENDIX{A}{A: The tachyon potential and string field redefinitions.}

Here we wish to discuss whether it is possible to make a field redefinition
of the string field tachyon such that the string action is brought to a form
where one could rule out the existence of a local minimum.
Even at the level of open string field theory this seems hard to achieve.
The tachyon potential is of the form $V\sim -\tau^2 + g \tau^3$. The cubic
term produces a local minimum with a nonzero vacuum expectation value
for $\tau$. We are
not allowed to just redefine $\tau$ to absorb the cubic term in the
quadratic one; this is
a non-invertible field redefinition. Using the massive fields is no help
since the transformations must preserve the kinetic terms, and therefore
should be of the form $\tau\to \tau+ f(\phi_i , \tau)$ and
$\phi_i \to \phi_i + g( \phi_i ,\tau )$, with $f$ and $g$ functions that
must start quadratic in the fields. Such transformations cannot eliminate
the cubic term in the tachyon potential.

Let us examine the question of field redefinitions in a more stringy way.
Assume it is possible to write the string action as
$$S = \int \d x \, \bigl[ \L (\nabla \tau , \phi_i) + \tau^2 (1+ f(\phi_i))
\bigr]\, , \eqn\tachleth$$
namely, that one can separate out a term just depending on derivatives of
the tachyon field, and all other fields, and a quadratic term for the
tachyon potential. The
term $f(\phi_i)$ was included to represent couplings to fields like
dilatons or background metric. If the above were true we would expect
no perturbatively stable minimum for the tachyon (the factor
$(1+ f(\phi_i))$ is expected to be nonvanishing). We will now argue
that the string action {\it cannot} be put in the form described
in Eqn.\tachleth\ by means of a string field redefinition. It is therefore
not possible to rule out a local minimum by such simple means.\foot{It is
not clear to us whether this result is in contradiction with that of
Ref.[\banks], where presumably the relevant action is the effective
action obtained after integrating out classically the massive fields.}

If Eqn.\tachleth\ holds, a constant infinitesimal shift of the tachyon
field $\tau \to \tau + \epsilon$ would have the effect of shifting the
action as
$$S \to S + 2\epsilon \int \d x \, \tau (1+ f(\phi_i)) + \O (\epsilon ^2)
\, .\eqn\tachses$$
We should be able to prove such ``low-energy tachyon theorem'' with
string field theory. For this we must find the change in the string
action as we shift the string field as follows
$$\ket{\Psi}_1 \to \ket{\Psi}_1  + \epsilon \ket{T_0}_1 + \epsilon
\bra{h^{(2)}_{23}}\Psi\rangle_2\ket{{\cal S}_{13}}+ \cdots \, ,\eqn\tryform$$
where $\ket{T_0}=c_1\bar c_1 \ket{{\bf 1}}$,
 the dots indicate quadratic and higher terms in the string field,
$\bra{h^{(2)}_{23}}$ is a symmetric bra, and $\ket{{\cal S}_{13}}$ is the
sewing ket [\senzwiebachtwo]. Indeed the transformation of the string field
cannot be expected to be a simple shift along the zero-momentum tachyon,
since the string field tachyon
should differ from the tachyon appearing in \tachleth. If we now vary the
string action \action\ we find
$$S \,\to \, S + \, \,\epsilon\bra{\Psi} c_0^-Q\ket{T_0}\, +{1\over 2}\epsilon
\left[ \, \bra{V^{(3)}_{123}}T_0\rangle_3 +\,
\bra{h^{(2)}_{12}}(Q_1+Q_2) \right] \ket{\Psi}_1\ket{\Psi}_2
+\cdots\,. \eqn\varacttr$$
We must now see that by a suitable choice of $\bra{h^{(2)}_{12}}$ the variation
of the action takes the form required by \tachses. Indeed, the term
$\epsilon\,\tau$ arises from the $\epsilon\bra{\Psi} c_0^-Q\ket{T_0}$
term in \varacttr\ since $ c_0^-Q\ket{T_0}$ can only couple to the tachyon
field in $\Psi$. Assume the function $f(\phi_i)$ in \tachses\ is  zero,
in that case there is no extra variation in the action, and we must
require that
$$ \bra{V^{(3)}_{123}}T_0\rangle_3 +\,
\bra{h^{(2)}_{12}}(Q_1+Q_2) = 0 \, .\eqn\musthave$$
This equation cannot have solutions; acting once more with $(Q_1+ Q_2)$ we
find that \musthave\ requires that
$\bra{V^{(3)}_{123}}Q_3\ket{T_0}_3 =0$ which cannot hold (recall
$Q\ket{T_0}\not= 0$). Even if $f(\phi_i)$ is not zero, we do not expect
solutions to exist. In this case we still must have that \musthave\ should
be zero contracted with any arbitrary two states, except when one of them is
a zero momentum tachyon. Again using states of the form
$(Q_1+Q_2)\ket{a}_1\ket{b}_1$ where neither $\ket{a}$ nor $\ket{b}$ is
a zero momentum tachyon, we see that again the equation cannot be satisfied.
This shows that there is no simple ``low-energy tachyon theorem'' that rules
out
a local minimum.

\refout
\end